\documentclass[twocolumn,showpacs,preprintnumbers,amsmath,amssymb,nofootinbib]
{revtex4-1}

\usepackage{graphicx}
\usepackage{dcolumn}
\usepackage{bm}
\usepackage{verbatim} 
\usepackage{epsfig}
\usepackage{amsmath}
\usepackage{amsthm}
\usepackage{subfigure}

\begin{document}

\date{\today}

\title{Frequency Domain Storage Ring Method for Electric Dipole Moment 
Measurement}
\author{Richard M Talman \\
Laboratory for Elementary Particle Physics,
Cornell University, Ithaca, NY, USA}

\begin{abstract}
Precise measurement of the electric dipole moments (EDM) of 
fundamental charged particles would provide a significant probe of physics
beyond the standard model. Any measurably large EDM would imply violation of 
both time reversal and parity conservation, with implications for the 
matter/anti-matter imbalance of the universe, not currently
understood within the standard model. A frequency domain (i.e. difference of 
frequencies) method is proposed for measuring the EDM of electrons or protons 
or, with modifications, deuterons. Anticipated precision (i.e. reproducibility)
is $10^{-30}\,$e-cm for the proton EDM, with comparable accuracy 
(i.e. including systematic error). This would be 
almost six orders of magnitude smaller than the 
present upper limit, and will provide a stringent test of the standard model.

Resonant polarimetry, made practical by the large polarized beam charge, is the 
key (most novel, least proven) element of the 
method. Along with the phase-locked, rolling polarization ``Koop spin wheel'', 
resonant
polarimetry measures beam polarization as amplitude rather than as intensity.
This permits all significant observables to be directly measureable as coherent 
frequencies. The same apparatus can be employed
to measure magnetic dipole moment (MDM) values with high accuracy. But this 
capability is more usefully exploited to determine ring parameters with
otherwise unachievable accuracy, using MDM values that are already known to 
high 
precision. Also novel, though less essential, is the M\"obius storage ring 
lattice 
modification, which greatly increases the spin coherence time (SCT), with
correspondingly improved accuracy. 

Important sources of EDM error, statistical or systematic, are 
considered, along with measures to be taken for improved accuracy.
Their effects can be expressed as the EDM upper limits they imply.
The polarization roll, at 100\,Hz for example, and adjustable by a 
constant control current, causes spurious torques due to
field errors to average to zero to high accuracy. Since these torques
have been considered to be the dominant source of systematic error in 
truly frozen spin operation, this is a major improvement resulting from the 
rolling polarization. Important sources of systematic errors remain, the
main one being due to Wien filter reversal uncertainty. 

Both electron and proton spins can be ``frozen'' in all-electric storage 
rings, and their EDM precisions should be comparable. Freezing the deuteron 
spin requires a superimposed electric and magnetic guide field; otherwise the 
rolling spin method and precision should be similar. But the deuteron 
option is not discussed in this paper.
\end{abstract}

\maketitle

\tableofcontents

\section{Introduction}
\subsection{Physics Justification and Current Status}
Quoting from the Final P5 Report\cite{P5}, of the Particle Physics 
Project Prioritization Panel (P5),
``Many extensions of the Standard
Model, including Supersymmetry, have additional sources of CP
non-conservation. Among the most powerful probes of new
physics that does not conserve CP are the electric dipole
moments (EDM’s) of the neutron, electron and proton.'' 

Various schemes for using frozen spin storage rings to measure
electric dipole moments (EDM) of various baryons have been suggested. 
An early design first proposed by Farley and others\cite{FFarley}
has, by now, evolved into a design by the International Storage
Ring EDM Collaboration\cite{BNLproposal}; important steps in this
evolution are described in various reports and papers.
They include: 
overall strategy, Semertzidis\cite{SemertzidisStrategy};
polarimetry, Stephenson\cite{StephensonPolarimetry};
squid magnetometry, Kawall\cite{KawallSquid};
magnetic shielding, Morse\cite{MorseMagnetic};
error compensation, Orlov\cite{Orlovcompensation};
lattice design, J. and R. Talman\cite{TalmanLattice}\cite{ETEAPOT1};
spin coherence time (SCT), Haciomeroglu\cite{Haciomeroglu} and Talmans\cite{ETEAPOT2}.
The main recent experimental advances toward the goal of storage
ring EDM measurement have occurred at the COSY storage ring
in Juelich, Germany, by Lehrach\cite{COSYnoMagLoss}, Rathmann, 
Stephenson and others, and described, for example, by 
Stroeher\cite{Stroeher}\cite{COSYpolarization}.

The storage ring EDM measurement proposed here has continued this 
evolution with major modifications. The most important modification
is the introduction of resonant polarimetry, which permits the EDM 
measurement to be ``moved into the frequency domain''. This, 
and other significant differences, are spelled out in detail in what follows. 
Much of the early discussion assumes electrons, but most results
are common to both electrons and protons; even deuterons in some
cases.

The current upper limit\cite{GabrielseElEDM}\cite{Regan} for the electron EDM 
is about 10 in our nominal $10^{-29}$\,e-cm EDM units. This measurement, using Thorium 
Oxide molecules, exploits a 9 order of magnification factor of internal to 
applied electric field for this molecule, helped also by a conveniently small 
molecular MDM. 

By constrast our storage ring trap measurement will measure the free electron EDM, 
using a much larger external electric field, but without benefit of the molecular
polarizability magnification factor. The expected instrumental precision (i.e. not 
including systematic errors) corresponds to an EDM value of $10^{-30}\,$e-cm. This 
is 100 times smaller than the current electron EDM upper limit, but with systematic 
errors not yet included. (For the proton the expected EDM precision is about the
same, and is almost 6 orders of magnitude smaller than the current proton 
upper limit.) 
It would not be realistic to claim, at this stage,
that all systematic errors at this level of precision have been
identified, much less eliminated. However, even if the $10^{-30}\,$ precision
is overly optimistic for ultimate accuracy, \emph{high precision is important, 
since it establishes the precision with which systematic errors can be 
investigated and ameliorated.} 

The most promising possibility, as regards physics reach, continues to 
be measuring the EDM of the \emph{proton} using an all-electric ring.
It is important to reverse the beam circulation direction frequently, 
with no ring modification (other than injection direction). This is required for
a significant reduction in systematic error. In previous proposals, such as
reference\cite{BNLproposal}, the two beams are required to counter-circulate 
simultaneously, as in a colliding beam (with ``collisions'' 
calculated to have negligible effect on the measurement).  
In the present proposal just one beam circulates at a time, but with 
frequent circulation reversals. The justification for dispensing with simultaneous 
beams is explained in connection with discussion of magnetic shielding, 
the purpose for which is to suppress the spurious EDM signal caused by 
poorly-known radial magnetic fields.

It has been realized only quite recently that charged hadron particle EDM's 
can be directly measured with accuracy comparable to or, actually, far better than, 
neutron EDM's. In particular there have been a continuing series of neutron
measurements over time, but no direct proton measurements. The history of
neutron EDM measurement, up to 1982, is described (and very clearly explained)
by Ramsey\cite{RamseyNeutronEDM}. Modern neutron EDM measurements use the
so-called ``Ramsey method of separated oscillatory fields''. This method is
discussed briefly in a later section on magnetic field shielding, needed
to reduce a source of systematic error common to neutron and charged particle 
EDM measurement. The analogy is much closer than magnetic shielding, 
however, since, like the proposed rolling spin method, the Ramsey neutron method 
can be regarded as shifting the EDM sensitivity ``into the frequency domain''.

\subsection{Proposed Method}
At their ``magic'' kinetic energies in an all-electric storage ring, 14.5\,MeV 
kinetic energy for electrons, 235\,MeV for protons, the beam polarization 
precesses at exactly the same rate as the beam momentum. When viewed at any 
fixed point in the ring, the polarization appears ``frozen'', for example 
always parallel to the beam orbit. By intentionally superimposing a
(precisely-controlled) torque around the local radial $x$-axis, the beam 
polarization will ``roll'' around that axis at frequency $f_{\rm roll}$. The
first of the two purposes for the roll is to make resonant polarimetry 
possible.

A resonant polarimeter responds directly to the longitudinal magnetization
of a polarized beam. With beam polarization frozen forward, the magnetization 
signal consists of harmonics of the revolution frequency. As such, the 
magetization signal would compete unfavorably with direct Coulomb excitation 
of the resonator by the beam charge, no matter how carefully 
the resonator is designed to be insensitive to that excitation. 
The rolling polarization alteration (first introduced by Koop, for 
different reasons, as a ``spin wheel'') overcomes this by shifting the 
magnetization signal frequency far enough away from revolution harmonics 
for the resonator quality factor $Q_{\rm res.}$ to be high enough to suppress the 
direct resonator response to the passing beam charge.  

With ideal stabilization and the magic condition exactly satisfied, the 
beam polarization stays always in the local $(y,z)$ plane where $y$ is 
``up'' and $z$ is longitudinal. In this condition, any torque due to the 
electric bending field acting on the electron EDM also lies in the $(y,z)$ plane 
and causes a systematic shift in the roll frequency. \emph{It is this frequency shift that is 
to be measured and ascribed to the electron EDM. Shifting the EDM 
sensing from polarimeter amplitude to polarimeter phase greatly
improves the ultimately-achievable EDM precision.} 

Both the longitudinal component of beam polarization vector and the roll
frequency will be measured using a longitudinally-aligned, resonant 
polarimeter, sensitive to the longitudinal beam polarization.  
As discussed in reference\cite{ResPolCornellWksp}, and in greater detail 
in a paper under preparation,
this polarimeter responds linearly to the slowly oscillating $z$-component 
of beam magnetization, and is insensitive to both radial $x$ and vertical 
$y$ polarization components, as well as being insensitive to 
beam charge. (Of course the magnetization signal \emph{is} proportional
to the beam charge. It is only because the polarized beam charge can be
large in a storage ring 
($10^{10}$ particles, more or less aligned, for example) that makes 
resonant polarimetry practical.) 

To keep the spin wheel properly aligned will require phase-locked loop
stabilization of the other two beam polarization degrees of freedom. 
Transverse components of beam polarization will be fed back to provide 
the required stabilizing 
torques. For protons this polarimetry can be performed by measuring 
both the 
left-right and up-down scattering asymmetries of extracted particles.
Alternatively, resonant polarimeters sensitive to transverse components of beam
polarization could be used. Such phase-locking has not yet been achieved in
practice. But the JEDI Collaboration\cite{COSYSpinTune} at the COSY ring in 
Juelich Germany, appears to be on the verge of performing this feat.

\begin{figure*}[ht]
\centering
\includegraphics[scale=0.4]{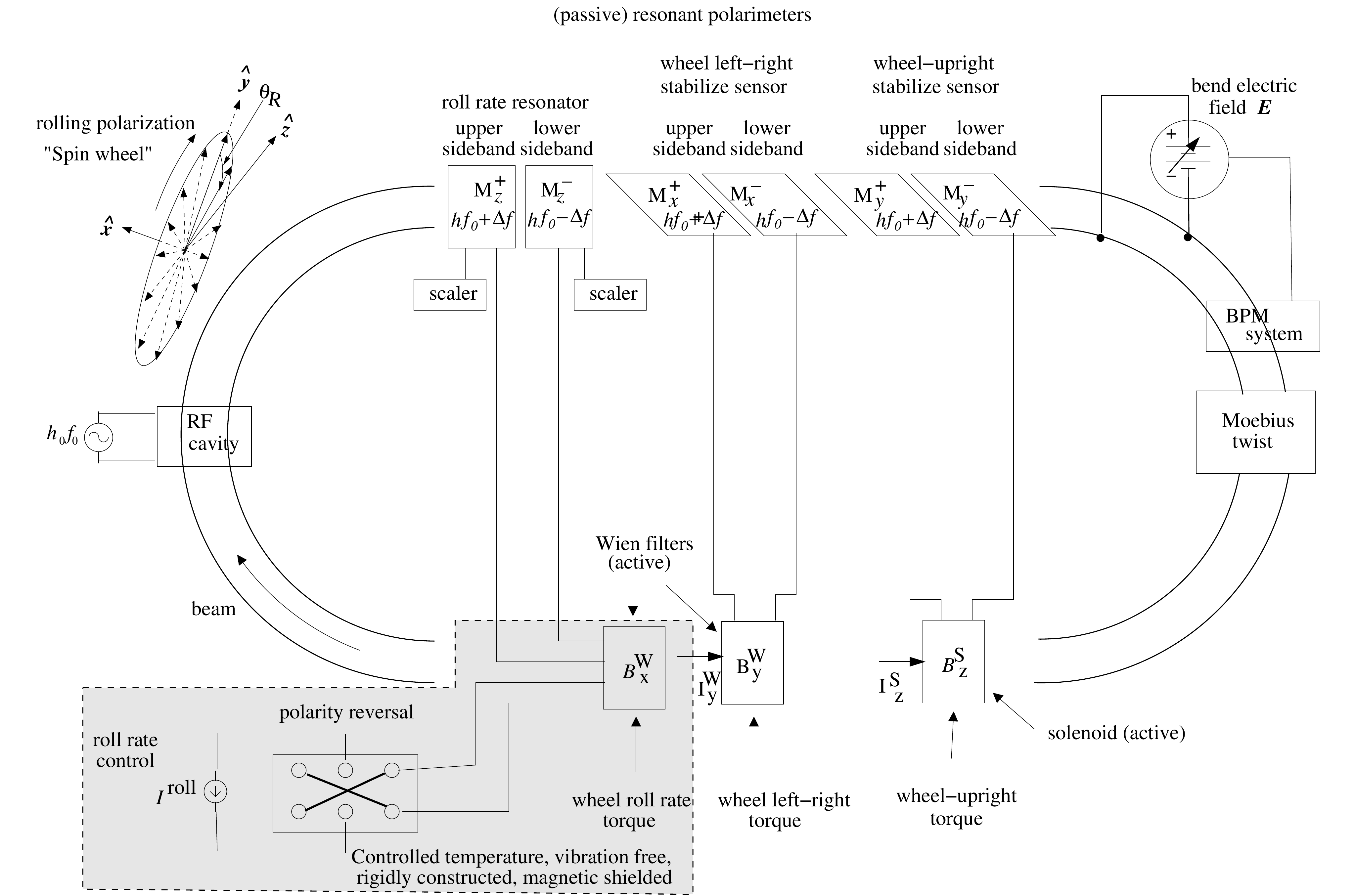}
\caption{\label{fig:SingleRing}Cartoon schematic of the ring
and its spin control. The Koop polarization ``wheel'' in the upper left
corner is to be visualized as rolling along the ring, always
upright, and aligned with the orbit. The boxes in the lower straight
section apply torques to the particle magnetic moments without
altering the design orbit. Elements with superscript ``W'' are
Wien filters; superscript ``S'' indicates solenoid. The frequency
domain EDM signal is the frequency change of spin wheel rotation 
when the $B^W_x$ Wien filter polarity is reversed. EDM measurement
accuracy (as contrasted with precision) is limited by the reversal
accuracy occurring in the shaded region. Precision is
governed by scaler precision.}
\end{figure*}

\begin{figure*}[ht]
\centering
\includegraphics[scale=0.6]{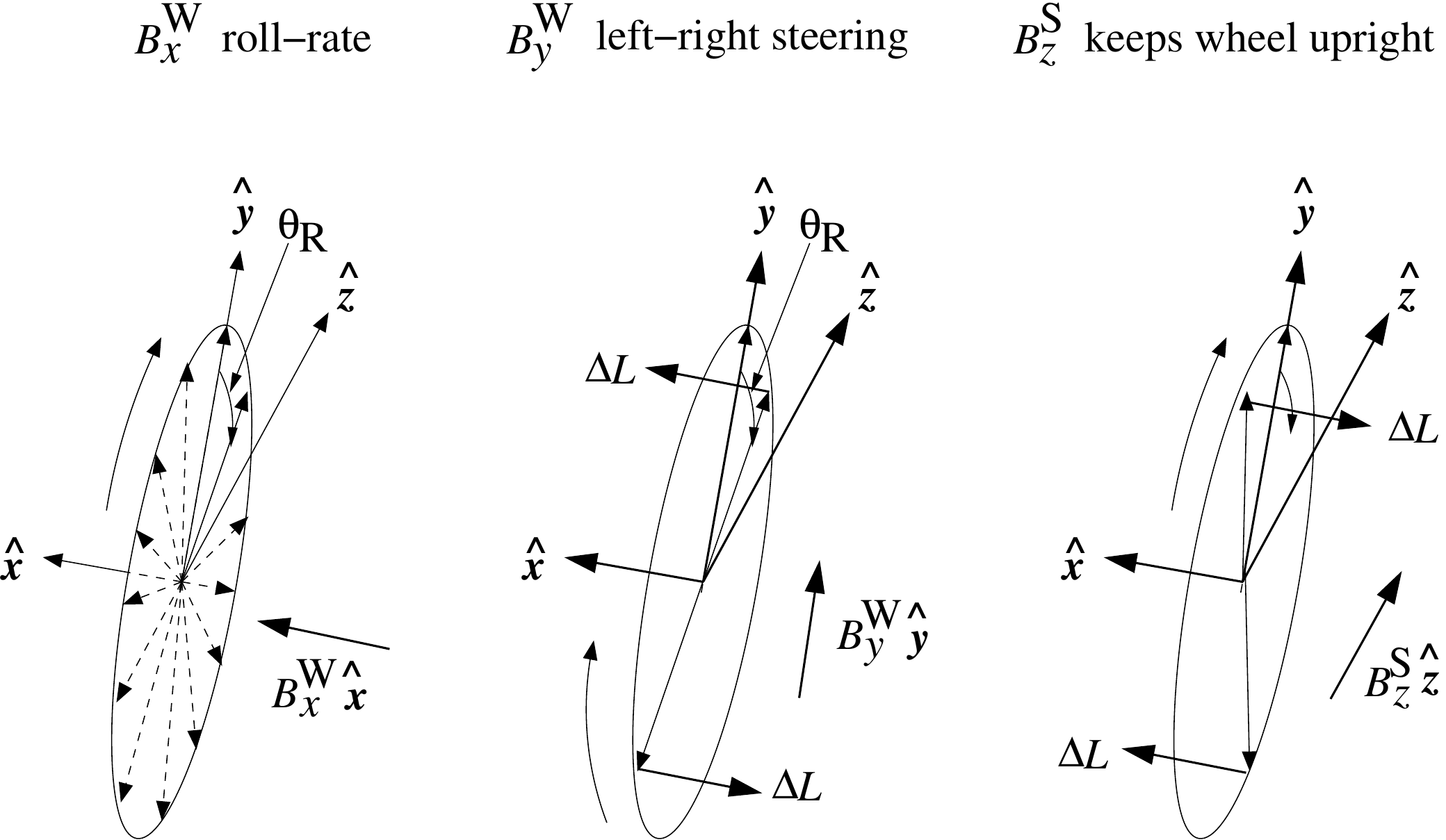}
\caption{\label{fig:By-out-steer}Roll-plane stabilizers:
Wien filter $B_x^W{\bf\hat x}$ adjusts the ``wheel'' roll rate,
Wien filter $B_y^W{\bf\hat y}$ steers the wheel left-right,
Solenoid $B_z^S{\bf\hat z}$ keeps the wheel upright.}
\end{figure*}

The proposed storage ring layout is shown in Figure~\ref{fig:SingleRing}.
Spin wheel, polarimeters, and stabilizers are shown.
Feedback to a Wien filter holds the roll plane perpendicular to the local
radial axis by providing left-right ``steering stabilization'' (though 
it is the polarization rather than the orbit that is being steered). 
This amounts to holding the average beam energy exactly on 
the magic value. The design lattice has no intentional solenoids. 
However a trim solenoid will be required to cancel possible solenoidal 
fringe field components. This solenoid also provides the 
``wheel-upright'' stabilization torque. The stabilizing torques are 
shown in Figure~\ref{fig:By-out-steer}.

Even if the proton EDM measurement is ultimately more promising 
as ``physics'', the electron 
EDM measurement is also important.  And it is sufficiently similar for
electron and proton measurements to be described together in this paper. 
Also, because the 
electron magic momentum is so much smaller, the electron ring will be 
much cheaper. A sensible first step might therefore be to first build, 
as a prototype, an all-electric, frozen-spin storage ring for electrons 
of kinetic energy $K=14.5\,$MeV (which is the ``magic energy'' at which the 
electron spin is ``frozen'' because the ``spin tune'' $Q^E_s$ vanishes). 
As well as serving as a prototype for an eventual proton ring, such an
electric ring could also be used to measure the electron EDM with
unprecedented precision, and to develop techniques applicable to the
proton measurement.

The spins in a beam of deuterons cannot be frozen in an all-electric
ring; superimposed electric and magnetic fields are required to cancel the
spin tune. The presence of magnetic bending excludes the possibility of 
simultaneously counter-circulating beams. Since 
counter-cirulating beams are not required to be simultaneous for the EDM 
measurement method proposed in the present paper, the same method will be 
applicable for deuterons, with the only added complication being that reversing 
the beam circulation direction will require reversing the magnetic fields.  
Nevertheless, for brevity, the deuteron measurement is not discussed 
in this paper, except for pursuing a suggestion by Senichev\cite{Senichev},
concerning the possibility of a deuteron ring with disjoint electric
and magnetic bend sectors. This is pursued later in the section titled
``Geometric Phase Errors''. 
 
I. A. Koop\cite{KoopSpinWheel} has proposed a ``spin wheel'' in which, by
applying a radial magnetic field, the beam polarization rolls in a
vertical plane with a frequency in the range from 0.1 to 1 Hz. The
rolling polarization I propose is similar, except my proposed roll 
frequency is greater by a factor as great as one thousand. 

As shown in the upper left corner of Figure~\ref{fig:SingleRing},
the beam polarization ``rolls'' at uniform
rate in the plane defined by the vertical $y$ and (local) longitudinal
$z$-axes. This rolling action causes the longitudinal polarization, 
which is what the polarimeter senses, 
to vary sinusoidally with frequency $f_{\rm roll}=\omega_{\rm roll}/(2\pi)$.
$f_{\rm roll}$ will be adjustable, to a value such as 100\,Hz. 
The ring revolution frequency is
$f_0=\omega_0/(2\pi)\overset{\rm e.g.}{\ =\ }10^7\,$Hz for electrons.
The polarimeter natural frequency will be tuned to one or
the other sidebands of harmonic number 
$h_r$ times the revolution frequency. 
\begin{equation}
f_{\rm res.\pm}=h_rf_0 \pm f_{\rm roll}
\label{eq:PolarimeterFreq.0}
\end{equation}
where the roll can be either ``forward'' or ``backward''.

The polarimeters 
have high $Q$-values, such as
$Q_{\rm res.}=10^7$ and fractional frequency selectivity of about
$1/Q_{\rm res.}=10^{-7}$. In theory the polarimeter is sensitive
primarily to beam magnetization and not to beam charge. 
However it is anticipated that, if tuned exactly to any harmonic of $f_0$, 
excitation due to direct beam charge or beam current would be likely 
to dominate the polarimeter response.  The rolling of the polarization 
shifts the polarization response frequency by an amount large compared
to the resonator selectivity. 

It is also essential for the rolling polarization to not
suppress the EDM signal. This is only possible if the
bunch polarization stays in a plane normal to the electric 
field, which is the source of the EDM torque. Since the
electric field is radial ($x$), the roll plane has to be 
vertical/longitudinal ($y,z$). The effect of the EDM torque is then 
to alter the rate of roll. The instrumentation has to distill 
this ``foreground'' effect from the ``background'' of intentional 
(and unintentional) sources of roll.

Just mentioned at this point, and discussed later in connection with
suppressing systematic errors, it can be mentioned that the 
rolling polarization will
also have the beneficial effect of helping torques due to field
errors to average to zero over times much less than one second. 

The ring is racetrack-shaped with most of the instrumentation in the long 
straights. ``Wien filters'' are crossed electric and magnetic fields which
cause no beam deflection. With no deflection in the laboratory
there is no electric field in the electron rest frame.  As a 
result these elements apply zero torque to the electron's EDM.
The horizontal and vertical Wien filter strengths are 
$B_x^W$ and $B_y^W$. In a Wien filter (stripline terminated by appropriate
resistor) such as $B_x^W$, both the electric and magnetic fields 
are produced by the same current, $I_x$.
\begin{equation}
B_x^W = I^W_xb_x^W.
\label{eq:PolarimeterFreq.0p}
\end{equation}
The electric field is produced by the voltage in the precision
terminating resistor carrying the current $I^W_x$.

The remaining stabilization field 
is $B_z^S$ which is the field of a solenoid aligned 
longitudinally. In all cases, though labelled
as if magnetic fields, it is actually the currents through
these elements that control the applied torques.
As with the Wien filters, the solenoid 
strength is established by a single current.

The beam magnetization observed at a fixed point in the ring
consists of a ``comb'' of periodic time domain impulses at the beam 
revolution frequency $f_0$, but with pulse amplitude varying 
sinusoidally at the roll frequency. The frequency domain 
representation of the magnetization then consists of upper and 
lower sidebands of the frequency domain comb of all harmonics of the 
revolution frequency. A resonant polarimeter sensitive to longitudinal 
polarization and with axis aligned along $z$ and tuned to frequency 
$f_{\rm res.\pm}=h_rf_0 \pm f_{\rm roll}$, where $h_r$ is 
an integer harmonic number, will respond to the upper or lower 
sideband excitation signal. Even when designed to be insensitive 
to direct excitation due to beam charge or beam current, a 
polarimeter tuned to any harmonic of the revolution frequency would 
likely be overwhelmed by direct Coulomb excitation. The roll 
frequency therefore has to be high enough, and the
quality factor $Q_{\rm res.}$ of the resonator large enough, to 
reject background response due directly to the beam charge or 
the beam current. This is what makes the rolling beam polarization 
essential for the EDM measurement.

\subsection{Definition of ``Nominal'' EDM} 
Only upper limits are presently known for the electric dipole
moments of fundamental particles. As regards physics, this is
unsatisfactory. As regards describing the apparatus to be
used in measuring EDM's it is a nuisance. To simplify
discussion it has become conventional to define a nominal
EDM value and define $\tilde d$ to be the EDM expressed in 
units of the nominal EDM.
 
Magnetic parameters are given for electrons and protons in 
Table~\ref{tbl:MagneticParams}. 
For an ideal, non-relativistic Dirac particle in a uniform magnetic field the 
momentum precession rate and the spin precession rate are equal---all energies 
are ``magic''. For a non-ideal particle with anomalous magnetic moment $G$ and 
magneton value $\mu=e\hbar/(2m)$, the difference between these precession rates 
is the anomalous precession rate $G\mu/\hbar$. By a numerical coincidence---though 
the electron's magnetic moment is three orders of magnitude greater than the 
proton's, its anomalous magnetic moment is three orders of magnitude 
less---the anomalous precession rates for electron and proton having the same
relativistic $\gamma$ factor are roughly the same. The same rough equality holds 
also with electric bending, with $B$ replaced by $E/v$. This provides a handy 
mnemonic when electron and proton EDM rings are being contemplated at the same time. 
Any precession in a storage ring that is due to a particle's EDM competes with
the anomalous precession due to its MDM, which is roughly the same for
electrons and protons. For assessing their relative importance, one can
compare the absolute precession due to its EDM with the anomalous precession 
due to its MDM.
\begin{table}[h] \scriptsize
\caption{\label{tbl:MagneticParams}e and p magnetic parameters. The units are SI, 
but with energies expressed in eV. Anomalous precession rates are in angular units 
of radians/second. In some cases the numbers are given to many-many places to indicate 
the accuracy to which the values are known, \emph{not} to indicate the values should 
be trusted as authoratively correct.}
\begin{tabular}{|c|c|c|c|c|}  \hline
   &  parameter            & symbol           &  value                     &  unit            \\ \hline	
e  &  Bohr magneton    & $\mu_B=e\hbar/(2m_e)$ & $5.7883818066\times10^{-5}$ &  eV/T            \\ 
   &  g-factor             & $g_e$            &  −2.00231930436182         &                  \\ 
   & anomalous mag. mom.   & $G_e=(|g_e|-2)/2$  &   0.0011596521809          &                  \\
   & Larmor prec. rate      & $g_e\mu_B/\hbar$ &  $-1.760859708\times10^{11}$ &  ${\rm s^{-1}/T}$  \\
   & anom. precession rate & $G_e\mu_B/\hbar$ & $1.019809775\times10^8$    &  ${\rm s^{-1}/T}$  \\ \hline
p  & nuclear magneton & $\mu_N=e\hbar/(2m_N)$ & $3.1524512550\times10^{-8}$ & ${\rm eV/T}$      \\ 
   &  g-factor           & $g_p=2\mu_p/\mu_N$ &   5.585694702              &                   \\
   & anomalous mag. mom.   & $G_p=(g_p-2)/2$  &   1.792847356              &                   \\
   & Larmor prec. rate      & $g_p\mu_N/\hbar$ &  $2.675222005\times10^8$        &  ${\rm s^{-1}/T}$  \\
   & anom. precession rate & $G_p\mu_N/\hbar$ &  $0.859\times10^8$         &  ${\rm s^{-1}/T}$  \\ \hline
\end{tabular}
\end{table}
For both protons and electrons we define a {\bf nominal EDM of $10^{-29}\,e$-cm}.
The EDM-induced precession rate (with $B$ replaced by $cE$ because
of the SI units) is then
\begin{equation}
\frac{d_{\rm nom}c}{\hbar}
=
\frac{10^{-29}\times(0.01)\times3\times10^8}{6.58\times10^{-16}}
=
4.56\times10^{-8}\,{\rm s^{-1}/T}. 
\label{eq:ratio.2}
\end{equation}
This is expressed in ${\rm s^{-1}/T}$ SI units, which are natural for expressing
precession rates of MDM's in magnetic fields measured in Tesla. 
Using a value from Table~\ref{tbl:MagneticParams}, 
we then determine the ratio of the EDM-induced precession in an electric field
to the MDM-induced anomalous precession in a magnetic field.
For electrons the relative-effectiveness ratio is
\begin{equation}
\eta^{(e)}_{\rm EM}
 = 
\frac{4.56\times10^{-8}}{1.0198\times10^8}
 =
0.46\times10^{-15}.
\label{eq:ratio.3}
\end{equation}
(From the table one sees that the corresponding ratio for protons 
is not very different.)
This ratio provides a semi-quantitative measure
of the relative difficulty of measuring EDM's compared to 
MDM's. Its smallness is what suggests that frequency domain methods 
will be required to isolate a statistically significant EDM 
signal.
 
``Frozen spin'' operation amounts to ``balancing on'' an unstable
equilibrium condition at an integer spin resonance. Such configurations 
are routinely investigated using
``Froissart-Stora'' scans\cite{Froissart} in which the spin tune is varied 
at a controlled rate, slow or fast, across the resonance. There is
secular precession and, eventually, for slow rate,
a complete flip of the polarization, for example from up to down,
as ring parameters are adjused
adiabatically from one side of the resonance to the other.
For the proposed EDM measurement this rate has to be made arbitrarily 
slow. If the electron EDM were huge it would dominate Froissart-Stora 
polarization reversals, making the EDM immediately measurable. 
But without great care, both electron and proton EDM's will be all but 
negligible compared 
to other effects capable of inducing Froissart-Stora polarization 
reversals. The primary discrimination comes from the mutual orthogonality
of the rest frame EDM and MDM pecessions.

For any realistically-small EDM the EDM-induced precession angle will not
exceed ten milliradians over runs of length comparable with spin coherence
time SCT.
Even with perfect resonance, this could not, by itself, cause even a single 
polarization reversal. The best that can be hoped for is a measurably 
large change over time in the orientation of the beam polarization due 
the EDM. To be able to extract 
such an EDM effect requires all other resonance drivers to
cause only exquisitely small ``wrong (for MDM) symmetry'' precession of bunch 
polarizations---certainly small compared to $\pi$, which would correspond 
to a complete spin flip. But, with polarization reversals and multiple 
run repetitions, the ``background'' precession can be 
smaller than the EDM-induced ``foreground'' precession

\subsection{The Frequency Domain ``Advantage''}
A feature of the proposed resonance polarimetry is that the EDM signal is 
encoded into a sinusoidal ``frequency domain'' signal. This signal can be digitized 
by counting cycles (``fringes'') or, with harmonic scaling, fractional cycles 
(``fractional fringes'' (FF)). Generally speaking, highest precision measurements
of physical constants rely on some version of this procedure. 

Before extolling the merits of the frequency domain, it is appropriate to consider
its disadvantages. The first step required to implement resonant polarimetry 
in the EDM measurement is to
add a large precession to the very small EDM that is to be measured.
Then, later, subtracting exactly the same precession to produce the small EDM
measurement. As arithmetic this is perfect but, as experimental physics, it
seems crazy. To the extent the added and subtracted \emph{large} signals
are not identical, when interpreted as a fractional error on the small signal,
the fractional error is magnified by the ratio of large to small signals. 
We are not talking about a small effect here. For typical parameters, using
a nominal $10^{-29}\,$e-cm EDM value for the small signal, the ratio of large 
to small roll rates will be about $10^{10}$ even for a small roll frequency
such as 100\,Hz. The added and subtracted integrated torques
providing the rolls have to be equal to better than this accuracy. 
Failure in achieving this will probably dominate the ultimate EDM
measurement error.

Reducing the roll rate improves the situation. If the roll-rate could
be reduced to zero---it cannot---the side band frequency displacement 
would give the EDM directly. The closest thing to this will be to measure
at two or more roll rates and extrapolate to the zero-roll point. 
Even though the direct measurement would be independent of the roll in 
this limit, the extrapolation from the data points will not be, and
the precision with which the roll reversal can be performed (on the average) 
will limit the ultimate precision. 

When counting whole cycles there is an unavoidable, $\pm 1$, least count, residual 
error. Consider measuring the frequency difference between two nearly identical
frequencies (for example a carrier frequency and one of its sidebands) using
two uncorrelated scalers. Without care, least count errors can introduce a 
hopelessly large error in the measured frequency difference. For better
precision the two scaler start and stop times have to be better synchronized
and fractional cycles recorded.

One way or another, when scaling sinusoids, the coherence of the quantities being 
measured has to be somehow exploited. This coherence cannot be reliably analysed 
without introducing the effects of noise.  In fact, it is noise rejection, rather 
than any digital/analog advantage, that may be the greatest frequency domain
advantage for the EDM measurement.

Suppose two sinusoids are known to have exactly the same frequency but unknown 
phase. As random variables, their phases are uniformly distributed on
the range from 0 to $2\pi$. Their phase difference (though perfectly constant)
is similarly uncertain.  One strategy for reducing the least count error would
to include simultaneous measurement of both phases. In an
EDM measurement of duration $10^3$ seconds this might proceed by measuring 
both phases 
during the initial $10^2$ second interval and again, later, 
during the final $10^2$ second interval.
The least count error is not, strictly speaking, stochastic but, pretending
it is, the averaging can be estimated to improve the average start and stop times 
proportional to the square root of the number of samples. This yields, for example, 
as the smallest detectable fraction of a cycle,
$\eta_{FF} = 1/\sqrt{10^2\times10^7}=0.00003$. A considerably larger, more
conservative, fractional fringe value, $\eta_{FF}=0.001$ is used in our later thermal 
noise error estimates. 

In the proposed method, the onus for carrying the EDM signal
has been shifted from polarimeter amplitude (i.e. a phasor quantity
whose squared magnitude is an intensity) 
to resonant polarimeter phase. To claim this will give increased 
precision one must first consider how the change alters the error analysis 
which produces the error estimate. To simplify the discussion
one can pretend there is no spin decoherence and there are no spurious 
signals whatsoever mimicking the EDM effect. For example the beam and 
the electric lattice are both 
perfect and there are never any magnetic fields whatsoever, nor time
variation of any sort. At the end
of a run of duration $T_{\rm run}$, some polarization angular displacement 
that is unambiguously due to the EDM will have developed. When this polarization
is measured by left-right scattering asymmetry there will be an inevitable,
unambiguous counting statistics error, depending, for example, on polarimeter 
analyzing power. 

Instead of scattering asymmetry, one could use the resonator\ \emph{amplitude}
(the square root of an intensity). Poorly-known resonator parameters, will 
make this error hard to estimate and undoubtedly unacceptably large. But 
the resonator response\ \emph{frequency} is not subject to this criticism.

The error is very different when the EDM effect is contained in the phase 
of a resonant polarimeter phasor amplitude. The only irreducible 
error would be due to the resonator ``phase noise'' accompanying the thermal 
excitation of the resonator. Even this ``irreducible'' error could, in practice,
be reduced; for example by reducing the temperature or the bandwidth of the 
resonator, or by using multiple resonators. 
At absolute zero there would be no thermal noise and the resonator phase 
determination would be limited only by timing limitations---such as 
insufficiently high scaler frequency in the digital hardware---giving insufficiently
fine time resolution. With modern day digital electronics no such limitation
is to be expected, even for the smallest imaginable EDM value. Accepting this,
the irreducible phase error will be given by the temperature of the resonator
that is actually in use.

\subsection{Tentative Parameters} 
Because of the preliminary nature of the proposed method, and because
this paper discusses 14.5\,MeV electrons and 235\,MeV protons more or less
interchangeably, some parameters cannot be specified to better than an order 
of magnitude or more. Revolution frequencies will be about $f_0=1\,$MHz for protons
or $f_0=10\,$MHz for electrons. The resonant polarimeter frequency will 
be higher by a harmonic number perhaps in the range from $h_r=1$ to $100$. 
The roll frequency may be $f_{\rm roll}=100\,$Hz; (preferably lower, but probably 
higher during set-up).  By comparison, at nominal EDM value 
of $10^{-29}\,$e-cm, expressed as a roll frequency, the EDM roll frequency
to be measured is about $f_{\rm EDM, nom.}=3\times 10^{-8}\,$Hz. (When expressed
as resonator phase advance, this rate is increased by harmonic number $h_r$,
possibly favoring large values of $h_r$.)

Assuming spin coherence time $SCT=1000\,$s has become 
somewhat conventional and has, to some extent, been justified in various reports
\cite{BNLproposal}\cite{COSYSpinTune}\cite{ETEAPOT1}\cite{ETEAPOT2}.
This makes it natural to adopt run duration $T_{\rm run}=1000\,$s.

There are various reasons for choosing the resonator quality factor $Q_{\rm res.}$ as
large as possible. But the resonator settling time $Q_{\rm res.}/f_r$ has to be very 
short compared to the run duration; for example $Q_{\rm res.}/f_r=T_{\rm run}/1000$.
This gives a $Q_{\rm res.}$ value in the $10^7$ to $10^8$ range. This is 
achievably conservative, though high enough to require superconducting 
conductor and cryogenic temperature. 

\section{Error Analysis Strategy}

\subsection{Storage Ring as ``Charged Particle Trap''}
The possibility of storing a large number, such as $10^{10}$,
of identically polarized particles makes a storage ring
an attractive charged particle ``trap''. But, compared to a table top 
trap, a storage ring is a quite complicated assemblage of many carefully, 
but imperfectly, aligned components, powered from not quite
identical sources.

For detecting and measuring the EDM of fundamental particles, 
much has been made of the difficulty imposed by the smallness of their EDM 
values relative to their MDM values. There is one respect, though, in which 
it is helpful for the MDM to be ``large''. It has made it possible for the MDM 
to have been measured to exquisitely high precision. (Here the phrase 
``exquisitely high'' is being commandeered temporarily as a technical term 
meaning ``can be taken to be exact''.) For present purposes the MDM is to be 
treated as exactly known. 

High enough beam polarization,
and long enough spin coherence time SCT, make it possible to ``freeze'' the
spins for long enough to attempt to measure the EDM. 
In this frozen state, the importance of some inevitable machine imperfections, 
that might otherwise be expected to dominate the errors, is greatly reduced.
Examples are beam energy spread and ring element positioning and alignment
uncertainties. (With the benefit of RF-imposed synchrotron oscillation 
stability) the average beam energy is fixed with the same 
exquisitely high accuracy with which the MDM is known. The polarization
vector serves as the needle of a perfect speedometer.
With the RF frequency 
also known to exquisite accuracy, the revolution period is similarly well
known. Then, irrespective of element locations and powering errors, the 
central orbit circumference is, if not perfectly known, at least very well known. (The
minor reservation expressed here is associated with the run-to-run
variability associated with possible beam emittance shifts.) An abbreviation
intended to encompass all of these considerations will be to refer to
the storage ring as a ``polarized beam trap''.

The precise beam energy determination can be checked occasionally to 
quite good precision using resonant depolarization\cite{DepolEnergy}.
Run to run consistency with relative accuracy of $10^{-7}$ can be
expected. Depending 
on BPM precisions the closed orbit beam orbit positions may be fixed 
to, perhaps, one micron accuracy\cite{LightSourcePrec}\cite{LightSourcePrec2}
\cite{LiberaBPM}\cite{PrecisionBPM}\cite{EBPM-XBPMcorrelation}
at each of the beam position monitors (BPM).

Irrespective of the ring circumference, with the beam speed fixed, and the
RF period fixed, the closed orbit circumference is known to be
constant to arbitrarily high precision. Initially $f_0$ will be dead-reckoned 
based on the magic velocity and the nominal ring circumference of the
design closed orbit, which is 
assumed to pass through all element design centers. 
The actual central closed orbit will not, in fact, pass through these design
centers. But this does not matter; the central closed orbit will automatically 
settle nearby. To the extent ring lattice elements drift, the bend electric field, 
and all steering elements will be adjusted to hold the orbit as constant 
as possible at all beam position monitors (BPM). These adjustments have to 
be sufficiently adiabatic for the polarization feedback circuits to stay 
locked. As long as this is satisfied the circumference will 
never vary\cite{CircumferencePrecision}.

This strategem reduces the importance of some sources of error, but without
eliminating them altogether. Of course one will build the EDM storage ring as
accurately as possible. But being a ``trap'' relieves the need for 
obsessively precise storage ring parameter specifications. For example, 
r.m.s. element position precision may tentatively be taken to be $0.1\,$mm, 
and alignment precision $0.1\,$milliradian. These are just plausible
``place holders'' in the present paper. A real storage ring design will 
require serious analysis and determination of tolerances like these. 

The lack of concern about element absolute positioning must not to be 
confused as lack of concern for BPM, orbit positioning precision, even 
assuming the ring has been tuned to be a perfect trap. The direct operational 
EDM measurement will depend on subtracting results from consecutive pairs of 
resonator measurements, knowing that only a
single control current has been reversed. Precision can be 
obtained by guaranteeing the forward and reversed beam orbits are
identical. With the only intentional change having been a single control
current, as few as two BPM's can confirm the symmetry. 
Better, the symmetry can be monitored as averages over 
\emph{every} BPM. Micron ($10^{-6}\,$m) precision has been achieved in, 
for example, light sources\cite{LightSourcePrec}\cite{LightSourcePrec2}.
Commercial BPM's, e.g. reference\cite{LiberaBPM}, advertise 
precision of $1\,\mu$m with temperature coefficient less than 
$1\,\mu$m/degree-Celsius. Achieving this level of precision 
expoits ultrastable circulating beams, such as will be available in
EDM rings. Ten or a hundred times better accuracy has been claimed 
for the International Linear Collider collision 
point optics\cite{PrecisionBPM}, even without the benefit of a stable 
circulating beam.

\subsection{Categorization of Error Sources}
To estimate the precision with which an EDM can be measured, one must first 
attempt to identify all possible sources of error, expecting, at least, to
identify the most important ones.  

As mentioned already, with resonant polarimetry the estimation of 
achievable precision in measuring EDM's can usefully be separated into 
two parts: that due to thermal phase noise in the resonant polarimeter and 
``everything else''. Then to make progress, the latter has to be broken into 
parts to be investigated individually. 

Because the EDM is so small for all fundamental particles, there is
another useful separation of error souces.
Errors can be associated with the \emph{absolute} smallness of the EDM, or with 
the smallness of the EDM \emph{relative} to the MDM. In conjunction with even
very small field errors, the MDM is capable of producing spurious precession, 
indistinguishable from that due to the EDM.

The \emph{absolute} smallness issue can be associated
with the thermal resonator noise. This source of error is unavoidable and irreducible 
and is, therefore, at least in principle, the leading source of error of this type. 
Because other sources of phase noise would be subject to similar analysis the 
errors they cause should, preferably, be discussed in the same context. But some
are too uncertain to allow this.

Once thought to be the dominant source of EDM measurement error,
is failure to distinguish true EDM-induced precession from spurious, wrong-plane, 
MDM-induced, precession. This can be referred to as a ``relative precession'' task.
In the following subsection, as one component of ``everything else'', the error 
caused by unknown radial component of magnetic field will be associated with the 
smallness of EDM \emph{relative} to MDM.

Minimizing the EDM error will involve vast numbers of phase reversals.  
To estimate these errors it is necessary to analyse the precision with which the 
reversals can be performed.

An important distinction can be made between ``internal'' and ``external'' 
sources of electric or magnetic field errors. External errors are due to equipment
over which one has no control other than shielding or filtering. 
The most serious example is magnetic noise due to unstable power, 
to power lines, to passing vehicles, to transients associated with equipment being
turned on or off, etc. Internal errors are due to imperfection in the elements
of the ring itself, especially powered elements, for example because of leakage
currents. By and large external errors are more to be feared than internal errors.
This is because internal sources can be investigated in controlled ways and, perhaps, 
eliminated.  A huge
benefit of the rolling polarization is that it reduces the serious time-varying (AC)
magnetic field external error sources in exchange for introducing the internal
error source of uncertain roll-reversal balance. Nevertheless, many of the
problems due to field errors are common to both the frozen spin method and the
rolling spin method. So much of the subsequent discussion is common to both.

Another important distinction can be made between DC, or nearly DC, errors, and
AC errors. It is very hard to protect against external AC magnetic field errors,
but easy to shield against AC electric fields.

\subsection{Resonator Noise and ``Everything Else''}
No matter what the source of EDM error, with the
EDM value not known even approximately, for convenience, each source 
of EDM error can be expressed as the EDM upper limit it implies.  
In this paper resonant polarimetry is being emphasized and the EDM error 
can come either from thermal noise, or from ``everything else'', where the 
latter has to be broken down and addressed source by source.

An estimate givn previously, based on current technology, with all other 
error sources turned off, in the presence of thermal noise, a proton EDM value 
of $10^{-30}\,$e-cm would yield a statistically significant EDM signal in one 
year of running. This said, and though said to be unambiguous, it has to be 
confessed that this error estimate is still somewhat arbitrary. In the 
(unlikely) event that the error from ``everything else'' can be reduced below 
$10^{-30}\,$e-cm, then there will be ways of reducing the thermal noise further. 
For example, the temperature can be reduced. Or, more economically, multiple 
resonators can be employed, and their coherence exploited. While discussing 
precision timing, Kramer and Klische\cite{KramerKlische} discuss exploiting 
the coherence of multiple 
detectors. This would exploit the absence of noise correlations between 
thermal noise signals in separate resonators.

It is more likely that the EDM upper limit from ``everything else'' will 
exceed the thermal noise limit. If true this would still not make the thermal noise 
specification based on $10^{-30}\,$e-cm unnecessarily aggressive. As already
stated, high data precision facilitates the investigation and reduction of
systematic errors.  

\section{Resonant Polarimetry}
\subsection{Polarimetry Possibilities}
Various polarimetry strategies have been shown to be effective for proton
EDM measurement. A scheme using proton-carbon elastic scattering has been 
effective\cite{EdStevenson}. With counter-circulating beams, or
with gas jet target, p-p scattering is another possibility\cite{ResPol}.
Both of these measure transverse polarization. 
In an earlier note\cite{RT-NovelPol} I had introduced the resonant 
polarimetry which is the basis for this paper. The resonator responds to 
longitudinal bunch polarization and can therefore be used to measure 
longitudinal polarization. The fact that the electron MDM is greater than 
the proton MDM by a factor roughly equal the ratio of their masses makes 
resonant polarimetry easier for electrons than for protons. On the other hand, 
electron-atom colliding beam scattering polarimetry is less promising for electrons 
than is proton-carbon scattering is for protons. These statements are explained 
more fully in various publications.

Torques due to MDM's are huge compared to any
achievable electric dipole moment induced torque. Any scheme to
measure an EDM will have to exploit ``deviation from null'' signal
detection. That is, the configuration has to be arranged such that 
(ideally) the MDM causes zero signal in a channel in which the EDM 
gives a measurably-large signal.  For example, for frozen-spin protons 
in an all-electric lattice, with spin frozen forward, there is no 
intentional induced vertical polarization component except
that caused by the EDM. In the absence of systematic error 
any measurably-large vertical polarization accumulation is then ascribed 
to the proton EDM.  Proton-carbon or p-p polarimetry, because
they are sensitive to transverse polarization, can be used for this
measurement. Such scattering polarimeters are ``fast'', meaning 
they measure the polarization of every circulating bunch. But their precision 
is subject to unfavorable counting statistics, especially because any 
realistic asymmetry is proportional to the difference of nearly equal 
counting rates.

Though continuing to keep scattering polarimetry available for
some purposes, such as controlling transverse polarization,
this paper concentrates on resonant longitudinal polarimetry. 
Relying on build up over many turns, a resonant polarimeter 
is ``slow''---incapable of resolving individual bunches. 
The resonator can only measure the net polarization of
whatever bunches there are, circulating CW and/or CCW.

Resonant polarimetry has two substantial advantages over scattering 
polarimeters. One obvious advantage is that the polarimeter is 
non-destructive; it does not attenuate the beam. But a more significant 
advantage is that the resonant polarimeter sums phasor amplitudes while 
scattering polarimeters sum (or rather subtract) intensities 
in the form of counting rates. The precision of left-right or 
up-down scattering asymmetries are necessarily limited by counting 
statistic errors on two approximately equal rates that have to 
be subtracted.

The resonant longitudinal polarimeter has no such problem.
there is zero response to unpolarized beams.  (As explained
previously, bunch electric fields have the wrong frequency to 
excite the resonator.) 

\subsection{Brief Description}
The proposed EDM measurement relies critically on resonant polarimetry.
Based entirely on classical electrodynamics, theory predicts a passive
and completely deterministic signal proportional to the beam polarization,
not unlike signals from ordinary beam position or beam current monitors. 
To the extent such a signal is free of noise it will permit vastly more 
sensitive EDM detection than would be possible otherwise.

It is unclear who first proposed resonant polarimetry, which is 
based on the direct measurement of beam magnetization. The first clear 
analysis was due to Derbenev in 
1993\cite{DerbenevResonant1}\cite{DerbenevResonant2}\cite{DerbenevResonant3}.
In 2012 (independently, to my best recollection) I produced a somewhat more 
sophomoric, but largely equivalent, proposal and analysis\cite{RT-NovelPol}. 

Regrettably resonant polarimetry has not, as yet, been succesfully demonstrated 
in the loaboratory.

My analysis of resonant polarimetry, now developed in greater detail in a
paper in preparation, differs markedly from Derbenev's, but is consistent as 
regards expected signal levels. The two approaches stress different applications 
as being most promising. Derbenev proposes transverse polarization measurement 
at high 
energy in a conventional pill-box cavity, and gives detailed numerical examples.
I propose longitudinal polarization at low energies (as needed for a frozen 
spin ring) using a helical resonator optimized for the polarimetry application.

\begin{figure*}[ht]
\centering
\includegraphics[scale=0.25]{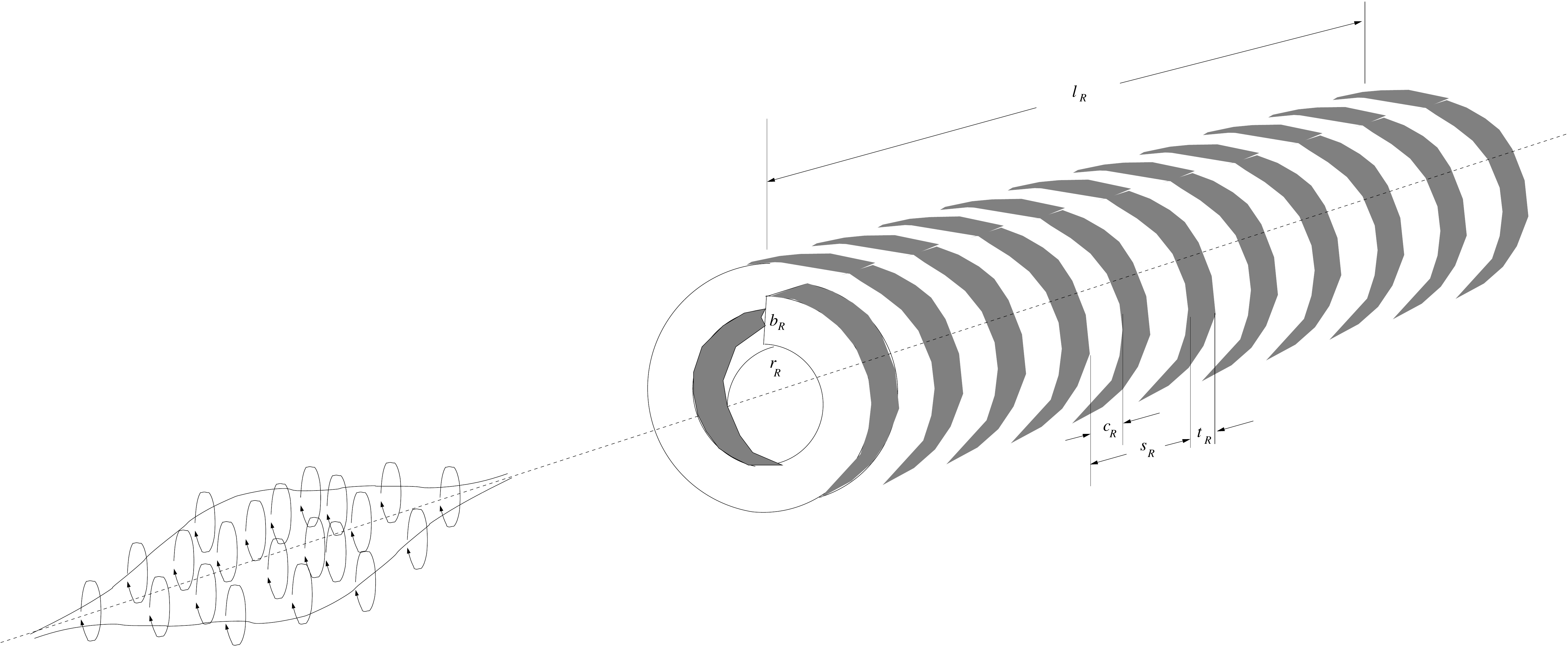}
\caption{\label{fig:HelicalResonator}Longitudinally polarized beam approaching
a helical resonator.
Beam polarization is due to the more or less parallel
alignment of the individual particle spins, indicated here as tiny 
current loops.}
\end{figure*}

My beam-resonator system is sketched in Figure~\ref{fig:HelicalResonator}.
The helical coil (inside a conducting cylinder not shown) acts as the
central conductor of a helical delay line. Depending on the particle speed
(which is close to the speed of light $c$ for electrons, but $0.6c$ for
protons, the wave speed and length of the transmission line are arranged
to meet two conditions. First, the fundamental resonance (or a harmonic)
of the open-at-both-ends transmission line, is tuned to a rolling polarization
sideband. Second, the relative values of particle and wave speeds through
the transmission line are such
that the resonator phase as the beam bunch exits the resonator 
is opposite to its phase on entry. 
This maximizes the energy transfer from beam to resonator, which maximizes 
the resonant response to a passing beam bunch. The work done by the
longitudinal Stern-Gerlach force is responsible for the energy transfer.
The bunch length has to be
short enough for the response to be constructive for all particles.

\begin{figure*}[ht]
\centering
\includegraphics[scale=0.55]{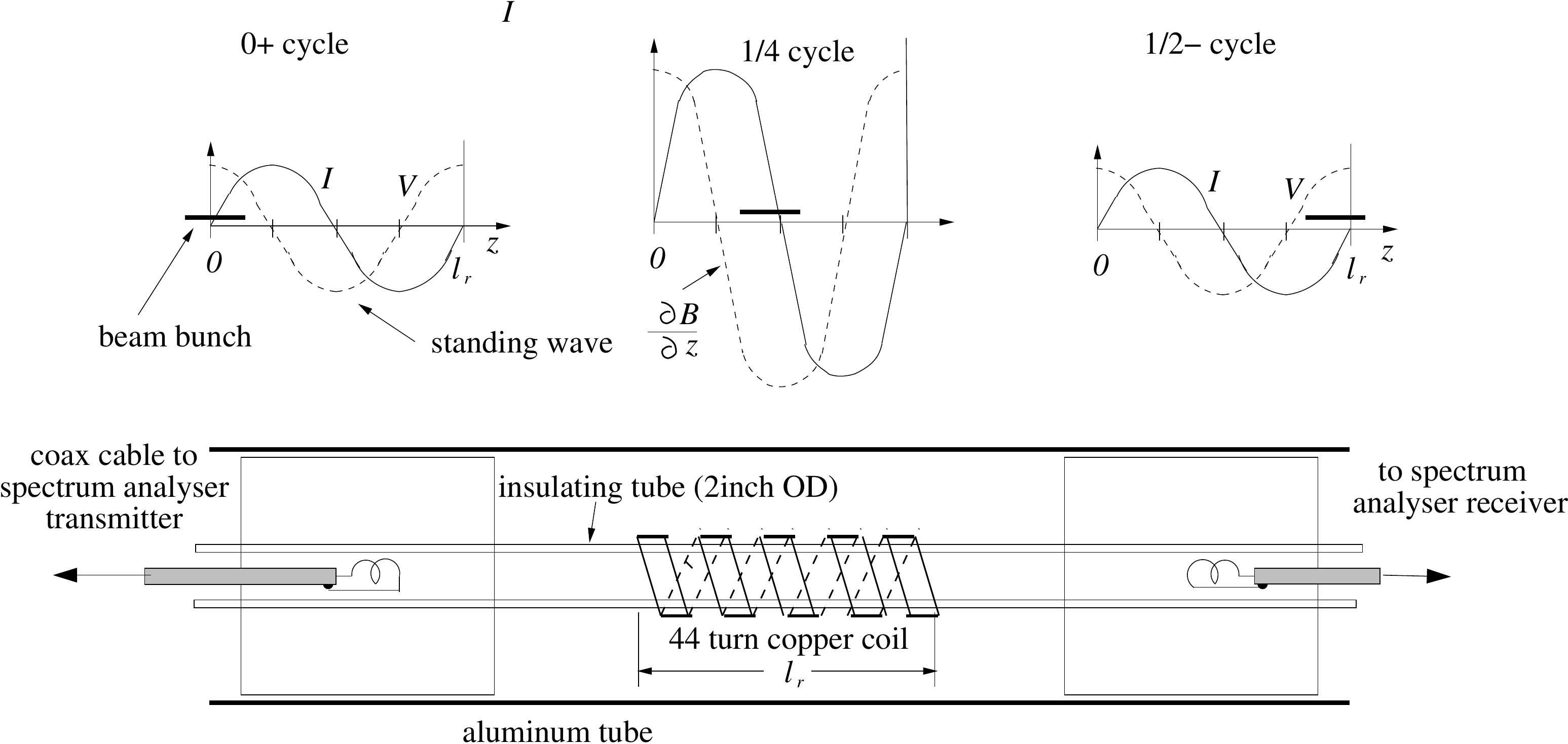}
\caption{\label{fig:HelicalResonatorTestRig}. Bench test set-up of 
prototype resonant polarimeter, with results shown in 
Figure~\ref{fig:MultiResSpectrum}. The coil length is $l_r$=11 inches.
Beam magnetization is emulated by the
spectrum analyser transmitter. Resonator excitation is detected by a single
turn loop connected to the spectrum analyser receiver. This would be an
appropriate pick-up in the true polarimetry application though, like the
resonator, the preamplifier 
would have to be at cryogenic temperature\cite{Huan}
to maximize the signal to noise ratio.
The figures above the apparatus are intended to complete the analogy to a 
situation in which the transmitter is replaced by the passage of a beam bunch.
The particle and wave speeds are arranged to maximize the energy transfer
from beam to resonator.}
\end{figure*}

\begin{figure*}[ht]
\centering
\includegraphics[scale=1.0]{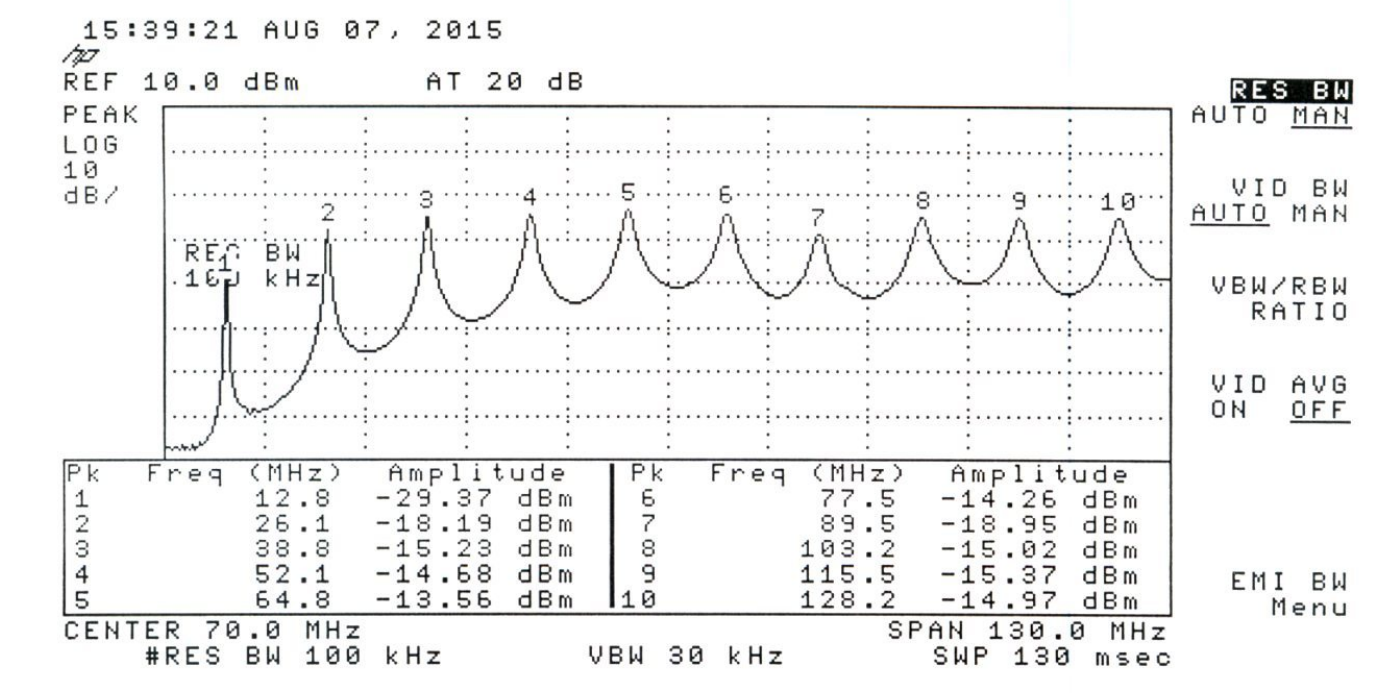}
\caption{\label{fig:MultiResSpectrum}Frequency spectrum observed using
the bench test shown in Figure~\ref{fig:HelicalResonatorTestRig}. Ten normal
modes of the helical transmission line are visible. The figure is cut off to
hide higher frequency modes, distorted by reflections in the
input or output cables. The systematic baseline shift may
be due to higher frequency modes riding on the tails of lower frequency modes.
Within measurement errors all resonator quality factors are about 200.}
\end{figure*}

A prototype resonant polarimeter, bench test set-up is shown in
Figure~\ref{fig:HelicalResonatorTestRig} and the spectrum analyser
output is shown in Figure~\ref{fig:MultiResSpectrum}. 
As expected, multiple (in this case 10) resonances are visible with
the frequency of the lowest being 12.8\,MHz, and
the others being integer multiples, as indicated in the table at the
bottom of the figure. Resonance quality factors are $Q_{\rm res.}\approx200$
for all of the resonances. Except for too-low $Q_{\rm res.}$-value, any one of
these resonance could serve for resonant polarimetry. For that matter,
any number of such coils (with frequency trimming not shown) could be tuned,
one each, to arbitrary sidebands of any harmonic $h_r=1,2,3,\dots10$ 
of the ring circulation frequency. 

Parameters for various possible polarimetry test configurations and
for sample electron and proton EDM rings are given in 
Table~\ref{tbl:SignalToNoise}. Full explanation of all entries in this
table would be too lengthy for inclusion in this paper. But the entries
in the first column are intended to be self-explanatory.
The bottom row, giving signal to noise ratios, makes optimistic assumptions about
achievable values of $Q_{\rm res.}$ and about noise reduction using coherent 
detection techniques.

\begin{table*}[h]
\caption{\label{tbl:SignalToNoise}Signal level and signal to noise ratio for various 
applications. In all cases the polarization is taken to be 1. In spite of the quite 
high $Q_c$ values achievable with HTS, the economy this promises is overwhelmed by 
the signal to noise benefit in running at far lower liquid helium temperature. 
The bottom row is the most important.
}
\medskip         
\begin{tabular}{|c|c|c|c|c|c|c|c|c|}  \hline
experiment          &         &       &   TEST   &  TEST    &  e-EDM   &  e-EDM         &  p-EDM   & TEST    \\ \hline
parameter           & symbol  &  unit & electron & electron & electron & electron       & proton   & proton  \\
beam                &         &       & J-LAB linac & J-LAB linac & ring &  ring        &  ring    &  COSY   \\ 
conductor           &         &       &  HTS     &  SC      &  HTS     &   SC           &   SC     &  SC     \\ \hline
ring frequency      & $f_0$   &  MHz  &          &          &   10     &   10           &   1      &  9.804  \\
magnetic moment     & $\mu_p$ &  eV/T & 0.58e-4  & 0.58e-4  & 0.58e-4  & 0.58e-4        & 0.88e-7  & 0.88e-7 \\ 
magic $\beta$       & $\beta_p$ &    &  1.0     &  1.0     &  1.0     &  1.0           &  0.60    &  0.6    \\ \hline
resonator frequency & $f_r$   &  MHz  &  190     &  190     &  190     &  190           &  114     &  114    \\
resonator radius    & $r_r$   &  cm   &   0.5    &  0.5     &   2      &    2           &    2     &   2     \\
resonator length    & $l_r$   &  m    &  1.07    &  1.07    & 1.07     &  1.07          &  1.80    &  1.80   \\
temperature         &  $T$    & ${}^{\circ}$K & 77 &   1     &   77     &    1           &    1     &  4      \\
phase velocity/$c$  & $\beta_r$ &   &  0.68    &  0.68    &  0.68    &  0.68          &  0.408   & 0.408   \\ 
quality factor      & $Q_{\rm res.}$   &       &  1e6     &  1e8     &  1e6     &  1e8           &  1e8     & 1e6     \\ 
response time       & $Q_{\rm res.}/f_r$ &  s  &          &  0.53    &  0.0052  &  0.52          &  0.88    & 0.0088  \\ \hline
beam current        &   $I$   &   A   &  0.001   &  0.001   &  0.02    &  0.02          &  0.002   & 0.001   \\
bunches/ring        & $N_b$   &       &          &          &  19      &  19            &  114     & 116     \\
particles           & $N_e$   &       &          &          &  1.2e10  & 1.2e10         & 1.2e10   & 6.4e9   \\ 
particles/bunch     & $N_e/N_b$ &     &  3.3e7   & 3.3e7    &  0.63e9  & 0.63e9         & 1.1e8    & 5.5e7   \\ \hline
magnetic field      & $H_r$   & Henry &  2.6e-7  & 2.6e-6   &  1.3e-6  &  1.3e-4        & 2.3e-6   & 1.15e-8 \\
resonator current   & $I_r$   &  A    &  2.2e-8  & 2.2e-6   &  2.2e-7  &  2.2e-5        & 0.50e-6  & 2.6e-9  \\
magnetic induction  & $B_r$   &  T    &  3.3e-13 & 3.3e-11  &  1.6e-12       & 1.6e-10  & 2.8e-12  & 1.4e-19 \\
max. resonator energy & $U_r$   &  J    &  2.9e-23 & 2.9e-19  &  2.9e-21        & 2.9e-17 & 1.5e-20  & 3.8e-25 \\
noise energy        & $\overline{U_m}$ & J & 0.53e-21 & 0.69e-23 & 0.53e-21  & 0.69e-23 & 0.69e-23 & 2.8e-23 \\ \hline
S/N(ampl.)          & $\sqrt{U_r/\overline{U_m}}$       & & 0.23  & 205 & 2.3 & 2055    & 45.8     &  0.117  \\
S/N(ph-lock)        & (S/N)$\sqrt{f_0}$ & ${\rm s}^{-1/2}$ & 3.2e3 & 2.8e6 & 3.2e4 & 2.8e7 & 4.9e5  &  1248   \\
\hline
\end{tabular}
\end{table*}

\clearpage

The resonant polarimeter theoretical analysis seems solid. Nevertheless, 
the method has not yet proven out experimentally.

Obviously the lack of experimental experience with resonant polarimetry 
introduces major uncertainty into the design of an experiment for which it 
is the main component. One thing is certain though; the signal 
level, relying as it does on Stern-Gerlach force, will be small. Both
Derbenev's and my paper calculate signal to noise ratios, assuming the noise
to be dominated by thermal noise at the resonator ambient
temperature. Both conclude that high-Q, cryogenic, superconducting resonators
are needed to bring the signal convincingly out of the noise. Certainly
stored polarized beams with as many as $10^9$ particles should give clean 
resonator output signals.

\subsection{Noise Limited Precision}
For the electron EDM experiment the predicted signal level is very safely
above the thermal noise level. This is also true for the proton EDM experiment,
though less so because of the smaller proton MDM. This
section analyses the influence of the noise on the polarimetry and produces
an estimate of the resulting precision limit.

Consider a particle circulating in a storage ring magnetic field with rotation
frequency $f_0$. Applying the $\eta^{(e)}_{\rm EM}$ ratio calculated previously,
the EDM-induced precession frequency is $\eta^{(e)}_{\rm EM}f_0$.
With revolution frequency of $f_0=$10\,Mhz, this is $6.283\times10^7$\,r/s.
0ne has to plan on measuring a ``nominal'' EDM-induced precession of order 
$3\times10^{-8}\,$r/s, or about 3\,mr/day. As a vector angle to be measured
this seems pretty small but, as a clock phase angle in a modern, high precision 
clock, it is not so small.

A virtue of frequency is that its measurement proceeds by counting 
cycles, or even fractions of cycles; for example $\eta_{\rm fringe}=0.001$, 
if one tenth percent of a cycle can be distinguished. 
Here a cycle is being 
referred to as a ``fringe'' as that term is applied in optical 
length measurements. It will be convenient for data analysis to 
allow the number of fractional fringes $N_{FF}$ to be non-integer and to
interpret $\eta_{\rm fringe}$ as a statistical error parameter 
(such as r.m.s. value) made in determining $N_{FF}$. One can then 
refer to $N_{FF}$ as the digitised polarimeter phase advance, expressed 
in units of the statistical error parameter.

Modern technology has enabled extremely accurate frequency comparison,
for example as described for a commercial device by Kramer and 
Klische\cite{KramerKlische}. They use the ``Allan variance'' 
$\sigma_y(T_{\rm run})$ as a quantitative 
measure of the fractional frequency variance after measuring for time 
$T_{\rm run}$. For the multi-MHz range we are interested in, their
apparatus achieves $\sigma_y=2\times10^{-15}$ for 
runs of the duration $T_{\rm run}$=1000\,s we are assuming. 
They characterize this performance as 
``better than 0.2 degrees related to 10 MHz'', 
Though Allan variance $\sigma_y$ is not simply converible into our 
$\eta_{\rm fringe}$ parameter, this corresponds roughly to 
$\eta_{\rm fringe}=0.0005$. This suggests our choice of 
$\eta_{\rm fringe}$=0.001, for a numerical example, is conservative. 
This conservativism reflects 
our current lack of understanding of the phase noise that limits the 
achievable precision. Kramer and Klische\cite{KramerKlische} suggest using 
coherent detection for further noise reduction. This technique would 
be made available to us using the coherent responses of multiple polarimeters.

The resonant polarimeter frequency is 
necessarily quite close to a harmonic of the revolution
frequency $f_0$. In other words $f_{\rm res.}/f_0$ is approximately equal
to an integer, such as 10. The rate at which fractional EDM fringes
accrue in the polarimeter is then  
\begin{equation}
f_{FF} = \frac{\eta^{(e)}_{\rm EM}}{\eta_{\rm fringe}}\,h_rf_0.
\label{eq:ratio.3p}
\end{equation}
In a pair of runs, each of duration $T_{\rm run}$, the total number
of fractional fringe shifts, after time $2T_{\rm run}$, is 
\begin{equation}
N_{FF} = \frac{2\tilde{d}\eta^{(e)}_{\rm EM}}{\eta_{\rm fringe}}\,h_rf_0T_{\rm run}
      \quad 
\Big( 
\overset{\rm e.g.}{\ \approx\ } 
\tilde{d}\,\frac{10^{-15}\cdot 10 \cdot 10^7 \cdot 10^3}{10^{-3}}
\Big).
\label{eq:ratio.3pp}
\end{equation}
where $\tilde{d}$ is the electric dipole moment in units of $10^{-29}\,$e-cm,
and $2\eta^{(e)}\approx10^{-15}$.
In this numerical example, with $\tilde{d}=10$,  after 2000\,s, $\tilde{d}$
will have been measured with one unit of statistical precision.

By averaging over some number of runs, $N_{\rm runs}$, or by performing longer 
runs one will have measured $\tilde{d}$ with higher precision. Exactly how the 
precision scales with $N_{\rm runs}$ depends on how we interpret the
$\eta_{\rm fringe}$ parameter. The most optimistic assumption possible is
that there is no random run-to-run error whatsoever, in which case the
precision improves proportional to $N_{\rm runs}$. 

Realistically, one expects 
some random run-to-run error. We can choose to interpret $\eta_{\rm fringe}$
as defining the r.m.s. run-to-run uncertainty, in which case the precision 
improves only proportional to $\sqrt{N_{\rm runs}}$. This is a pessimistic
result. Even in this case it will remain necessary to determine $\eta_{\rm fringe}$ 
phenomenologically. As has been mentioned already, the dominant contributor 
to $\eta_{\rm fringe}$ may be irreducible phase noise in the resonant polarimeter. 

To properly reduce the ambiguity between optimistic and pessimistic approaches
it would be necessary to replace $\eta_{\rm fringe}$ by two independent
statistical parameters. But this will be quite difficult, for example
because the split will depend on the run length and other uncertainties.
To be conservative we continue by simply following the pessimistic route. 
Continuing our earlier numerical example, and expecting to make $10^4$ pairs of runs 
(over about one year) one will achieve a one sigma EDM precision of
$10^{-30}\,$e-cm.

\section{Spin Precession}
\subsection{Field Transformations}
The dominant fields in an electron storage ring are
radial lab frame electric $-E{\bf\hat x}$ and/or vertical lab
magnetic field $B{\bf\hat y}$. They give\cite{Jackson}
transverse electron rest frame field vectors ${\bf E'}$ and ${\bf B'}$,
and longitudinal electric and magnetic components $E'_z$ and $B'_z$,
all related by 
\begin{align}
{\bf E'} &= \gamma({\bf E} + {\bf \beta}\times c{\bf B}) 
          = -\gamma(E+\beta cB)\,{\bf\hat x} \label{eq:Lorentz.3a} \\
{\bf B'} &= \gamma({\bf B} - {\bf \beta}\times{\bf E}/c)
          = \gamma(B + \beta E/c)\,{\bf\hat y} \label{eq:Lorentz.3b}\\
E'_z &= E_z, \\
B'_z &= B_z.
\label{eq:Lorentz.4}
\end{align}

\subsection{MDM-Induced Precession in Electric Field}
A particle in its rest system, with angular momentum ${\bf s'}$ and magnetic 
dipole moment $g\mu{\bf s'}$, in magnetic field ${\bf B'}$, is subject to 
torque $g\mu{\bf s'}\times{\bf B'}$. Here $\mu=e\hbar/(2m)$ is the ``magneton'' 
value for a particle of that particular mass and charge. By Newton's angular 
equation, substituting from Eq.~(\ref{eq:Lorentz.3b}) to express the
fields in laboratory coordinates,
\begin{equation}
\frac{d{\bf s'}}{dt'}
 =
-g\mu{\bf B'}\times{\bf s'}
 =
-g\mu\gamma\Big(B + \beta\,\frac{E}{c}\Big)\,{\bf\hat y}\times{\bf s'} .
\label{eq:Newton.1}
\end{equation}
Customarily the rest frame angular momentum is represented
by ${\bf s}$ rather than by ${\bf s'}$ (because ${\bf s}$ is a
true 3-vector only in that frame) and the laboratory
time interval $dt=dt'\gamma$ is used instead of rest 
frame time interval $dt'$. Furthermore, the magnitude $|{\bf s}|$
is known to be constant. With the 
laboratory field being purely electric and purely radial,
the vertical component of ${\bf s}$ is conserved.
The normalized horizontal component ${\bf\hat s}$ satisfies
\begin{equation}
\frac{d{\bf\hat s}}{dt}
 =
-\frac{g}{2}\,2\mu\beta\,\frac{E}{c}\,{\bf\hat y}\times{\bf\hat s}.
\label{eq:Newton.3}
\end{equation}
As Jackson explains\cite{Jackson-11.166}, relativistic effects cause 
the electron axis to precess in the laboratory, irrespective of any static moments
the electron may have. This
Thomas precession causes the polarization vector to precess 
even if there is no torque acting on the magnetic or electric 
moments. This precession has to be allowed for when ascribing precession
to MDM's or EDM's. In our case, the beam direction advances 
uniformly, by $2\pi$ during one revolution period $T_{\rm lab}$ and
the Thomas precession term is
\begin{align}
\frac{\gamma-1}{v^2}\,
&
\Big(
\frac{d{\bf v}}{dt}\times{\bf v}
\Big)\times{\bf\hat s}  
  =
\frac{\gamma-1}{v^2}\,
\Big(
\frac{-{\bf\hat x}v^2}{r_0}\times{\bf v}
\Big)\times{\bf\hat s} \notag \\
 &=
\frac{\gamma-1}{r_0/v}\,
{\bf\hat y}\times{\bf\hat s}
 =
\Big(
1 - \frac{1}{\gamma}
\Big)\,
2\mu\,\frac{E/c}{\beta}\,
{\bf\hat y}\times{\bf\hat s}.
\label{eq:Newton.7}
\end{align}
Adding this term to the electric part on
the rhs of Eq.~(\ref{eq:Newton.1}),
\begin{equation}
\frac{d{\bf\hat s}}{dt}\Big|_{\rm MDM,E}
 =
2\mu
\Big(
-\frac{g}{2}
 +
\frac{\gamma}{\gamma+1}
\Big)\,
\beta\frac{E}{c}\,
{\bf\hat y}\times{\bf\hat s}
.
\label{eq:Newton.8}
\end{equation}
This agrees with Jackson's Eq.~(11.170).
We try a solution of the form
\begin{equation}
{\bf\hat s}_{\rm MDM,E}
 =
\big({\bf\hat Z}\,\cos Q_{\rm MDM,E}\frac{v}{r_0}t - {\bf\hat X}\,\sin Q_{\rm MDM,E}\frac{v}{r_0}t\big).
\label{eq:Newton.9}
\end{equation}
For circular motion in an electric field, $r_0E=\beta(pc/e)$,
which leads to
\begin{equation}
Q_{\rm MDM,E}
 = 
G\beta^2\gamma-\frac{1}{\gamma} + 1.
\label{eq:Newton.10}
\end{equation}
Accelerator spin physicists define the ``spin tune'' 
$Q_s$ in an electric field by
\begin{equation}
Q_s
\equiv
\frac{d\alpha}{d\theta}\Big|_E
 =
G\beta^2\gamma-\frac{1}{\gamma},
\label{eq:SpinPrecess.4m}
\end{equation}
where $\alpha$ is
the angle between spin vector and particle velocity.
The two definitions of ``spin tune'' have 
therefore been inconsistent. Unlike the
beam direction, which advances by $2\pi$ each turn, and
could therefore be described as having a ``tune'' value of 1, 
the spin tune is reckoned 
relative to the particle velocity rather than relative 
to a frame fixed in the laboratory. This accounts for 
the ``+1'' on the rhs of Eq.~(\ref{eq:Newton.10}).)
The content of this section has therefore
amounted to being a derivation of Eq.~(\ref{eq:SpinPrecess.4m}), 
for the spin tune in an electric ring.

\subsection{EDM-Induced Precession in Electric Field}
A particle at rest, with angular momentum 
${\bf s}$ and electric dipole moment $d\ {\bf s}$, in electric field 
${\bf E'}$, is subject to torque $d\ {\bf s}\times{\bf E'}$.
Using Eq.~(\ref{eq:Lorentz.3a}), if the laboratory field is purely
electric, the induced precession satisfies
\begin{equation}
\frac{d{\bf s}}{dt}\Big|_{\rm EDM,E}
 =
d\ {\bf s}\times{\bf E'}
 =
-d\ {\bf s}\times E\,{\bf\hat x} .
\label{eq:EDMprecssion.1}
\end{equation}
This precession is small enough to be treated as a perturbative
addition to otherwise-inexorable polarization evolution.

To calculate EDM-induced precession we can use evolution formulas 
to describe dependence on $\theta$, of the beam angle, which advances by
$2\pi$ every turn.  With $({\bf X,Y,Z})$ being Cartesian coordinates
fixed in the lab, setting $Q_{\rm MDM,E}=1$, as appropriate for frozen
spin motion, the Frenet and spin vectors advance as
\begin{align}
{\bf\hat x} &= {\bf\hat X}\,\cos\theta + {\bf\hat Z}\,\sin\theta, \notag \\
{\bf\hat s}
 &=
{\bf\hat Y}\cos\Theta
 + 
\big(
{\bf\hat Z}\cos\theta - {\bf\hat X}\sin\theta
\big)\sin\Theta.
\label{eq:EDMprecssion.2}
\end{align}
Here $\Theta$ is the polar angle of the polarization relative to
the vertical axis.
Substituting these on the right hand side of 
Eq.~(\ref{eq:EDMprecssion.1}), the evolution of ${\bf s}_{\rm EDM,E}$ 
is the driven response given by
\begin{align}
\frac{v}{r_0}\,&\frac{d{\bf\hat s}}{d\theta}\Big|_{\rm EDM,E}
 =  \label{eq:EDMprecssion.3}\\
& -d\ E\,
\Big(
\cos\Theta(-{\bf\hat Z}\cos\theta + {\bf\hat X}\sin\theta)
 +
\sin\Theta\big(\cos\theta\big){\bf\hat Y}
\Big)
\notag
\end{align}

\section{Conquering $\Delta B_r$ Field Errors}
\subsection{Qualitative Discussion}
The Achilles heel of storage ring frozen spin EDM measurement has, 
until now, been
spurious precession due to any non-zero radial magnetic field average
$\langle B_r\rangle$. Acting on each frozen particle's MDM, this
magnetic field applies torque that perfectly mimics the EDM effect.
Without rolling polarization, the whole determination of systematic
error boils down to finding the maximum spurious
EDM precession that can come from time-dependent variation of 
$\Delta B_r$.

Starting with a discussion of frozen spin operation, much of this
section is devoted to showing that rolling polarization largely
eliminates this source of systematic EDM error. 

In the BNL proposed\cite{BNLproposal} proton EDM experiment
this precession error is limited by precision BPM-measured
cancellation of the vertical displacement of simultaneously 
counter-circulating, exactly superimposed beams.
Unlike spin precession due to $\langle B_r\rangle$, which is 
independent of beam direction, there is a differential displacement 
of the counter-circulating beams, one up, one down. Cancelling this 
differential displacement cancels both the spurious MDM-induced 
precession and the average magnetic fields produced by the two beams,
not just on the beam centerline, but also nearby. 
Using ultra precise squid beam position monitors, located as near 
as possible to the beam orbits, this method is
explained by Kawall\cite{KawallSquid}. The produced 
signals are proportional to the vertical beam separation.
The effectiveness of this approach depends on the
coherent summing of field amplitudes in each one
of many BPM's, along with the ``null'' beam centering this enables.
According to that proposal, this procedure limits the systematic EDM 
accuracy from this source to about $\pm10^{-30}\,$e-cm 
after running for one year.

A goal for the present proposal has been to reach at least the same level
of accuracy, but without depending on simultaneously circulating
beams. The rolling polarization method renders this CW/CCW beam reversal 
less essential. The new method continues to use counter-circulating beams,
but they circulate consecutively, not concurrently. This produces many 
of the same cancelations, but lack of simultaneous beams is less 
effective for two
reasons. One is that the vertical beam displacement is being measured
as the small difference of two positions, each of which is being 
measured using lower precision BPM's (because of their larger 
required dynamic range). The other is that, to the extent the 
magnetic field errors depend on time, their cancelation is impaired 
by measuring them at separate times. 

To be able to claim comparable insensitivity to $\Delta B_r$ errors,
without depending on simultaneously circulating beams, we have to 
overcome this loss of EDM selectivity. A 
``selectivity factor, S.F.'' will be used to reduce 
tedious circumlocution in the following discussion. Small S.F. is
good, S.F.=0 is perfect.

For ultimate precision
one will, in any case, try to situate the ring away from 
unpredictable magnetic sources in a city. With active and
passive shielding, fields of several fT/$\sqrt{\rm Hz}$ 
(femto-Tesla per root Hertz) have been obtained in shielded 
rooms in city environments. Averaged over a 1000\,s run, this 
gives about $\pm 10^{-16}$\,T. It is essentially impossible for this 
field to be always radial in the storage ring. Averaging over the 
full ring is likely to give a reduction factor of perhaps 10.

The magnetic field error is in competition with the 
``magnetic equivalent'' of an electric field of 
roughly $5\times10^6\,$V/m. Dividing by $c$, this is 
equivalent to $0.016$\,T. The ratio 
$10^{-16}/0.016=0.6\times10^{-14}$
has to be compared to the relative effectiveness factor 
$\eta^e_{\rm EM}=0.46\times10^{-15}$. By this estimate the
spurious MDM signal is 10 times greater than the nominal EDM
signal. 

We have been copying, so far, from reference\cite{BNLproposal},
assuming the beam polarization to be truly frozen. This is no 
longer the case in the present proposal. Now the beam polarization 
rolls with a frequency of, say, $f_{\rm roll}=100\,$Hz. Even if this frequency 
were imposed by a control voltage oscillating at frequency
$f_{\rm roll}$, it would produce substantial reduction of
the spurious precession caused by $\Delta B_r$. But any
correlation between roll-torque and roll-phase would 
limit the EDM selectivity improvement. In fact, the polarization 
roll-frequency is ``autonomously'' self-generated from an applied 
DC control current $I^W_x$. There is no source
for the rolling polarization to be coherent with. 
As a result the spurious torque due to $\Delta B_r$ will truly average
to zero over times of order $1/f_{\rm roll}$=10\,milliseconds.
Furthermore, even if, by chance, there were accidental synchronism
at one roll frequency, it would not be present at another.

This rolling polarization improvement has not come without cost. 
In our proposed method, the EDM measurement comes from differencing 
forward and backward roll rates. Canceling 
the effect of external magnetic fields has come at the cost of introducing 
new possible sources of error via this subtraction. 
This will be addressed shortly.

The discussion so far is largely applicable also to 
time-dependent $B_r$ fields generated inside a shielded
room or tunnel. Current leakage or sparking would be candidate 
sources. Permanent magnetization would not be serious, but 
hysteretic or temperature dependent magnetization could be. 
Repeated precision measurements with the same circulation
direction for both beams will give reliable information 
concerning the time-dependence of these magnetic field errors. 
This information will indicate the extent to which the EDM error
can be reduced by averaging over multiple runs. At a minimum 
this sort of control experimentation will permit an objective 
determination of systematic error.

Another concern is that a sufficiently large magnetic disturbance 
could destroy the phase lock. Pulses of this magnitude seem not
to have been observed in recent polarization measurements with
deuterons in the COSY ring at Juelich\cite{COSYpolarization}\cite{COSYnoMagLoss}. 
If such a pulse is large
enough for one or more of the phase-locked loops to lose lock
the run would obviously have to be discontinued and discarded.

\subsection{Spin Wheel Advantage for Field Errors}
\emph{As well as spotting two typos in this section (which have been fixed) J\"org Pretz has questioned the treatment of Thomas
precession and pointed out that the averaging in Eq.~(\ref{eq:NTLO.9p}) 
(the main result in this section) is incorrect. More careful treatment
has introduced a factor $1/\gamma_m$ into 
Eq.~(\ref{eq:NTLO.9p}). This factor is unimportant for proton EDM
measurement, but important for the electron. The averaging error 
pointed out by Pretz has not
been corrected however. It is left as an easy exercise---to figure out
why the averaging is incorrect (for the radial magnetic field component
at issue)---and a hard exercise---what to do about it. The eventual
success of the EDM measurement will depend on arranging conditions
to cancel this source of spurious precession.}

\emph{In compensation for the reduction in precision acknowledged in the previous paragraph, I am pleased to
report a significant improvement in suppression of radial magnetic
field systmeatic error\cite{Self-magnetometerBottle}. By using
the self-magnetometry feature of  
an octupole focusing electric storage ring ``bottle'', the average
radial magnetic field error can be cancelled with accuracy $\pm3\times10^{-16}$\,T. }

For comparison purposes, we continue to develop formulas applicable 
to both frozen spin and rolling spin methods.

For an apparatus intended to measure EDM's, one has to be prepared to
suppress any MDM-induced precession that mimics EDM-induced precession.
The leading source of systematic error is (unintended) rest frame radial magnetic 
field $B'_x$ which, acting on the MDM, mimics the effect of radial electric field
acting on the EDM. It is only the rest frame magnetic field ${\bf B'}$
that causes MDM-induced precession, but both laboratory frame components, 
$\Delta B_x$ and $\Delta E_y$, contibute to ${\bf B'}$.
These are also the lab frame field components 
capable of causing the closed orbit to deviate from the ideal, horizontal, 
design plane. Because the guide field is radial electric, but with
possible error from not
quite vertical electrodes, the dominant field
error can be expected to be a vertical laboratory electric component 
$\Delta E_y$. Transformed to the electron rest frame, this contributes 
a radial rest frame magnetic component $\Delta B'_x$.

To the extent the beam \emph{does} move out of the horizontal plane,
the cross product in the transformation from lab to rest frame
can also produce a radial magnetic field. This I neglect, relying on
precise beam steering, and particle trap operation, and hoping 
that the tendency to average to zero will not be defeated by conspiring 
correlations. Substituting into Eqs.~(\ref{eq:Lorentz.3b}), 
the fields being retained are, for electrons
\begin{align}
{\bf E'} 
 &= -E\gamma{\bf\hat x} +
(\Delta E_y(\theta) \pm \beta c\Delta B_x(\theta))\gamma{\bf\hat y},
\label{eq:NTLO.1} \\
{\bf B'}
 &= 
(\Delta B_x(\theta) \pm \beta \Delta E_y(\theta)/c)\gamma{\bf\hat x},
\label{eq:NTLO.2}
\end{align}
where the error fields depend on $\theta$ which is the angular position
around the ring.

Here the upper of the $\pm$ and $\mp$ signs refer to a clockwise (CW)
and the lower to a counter-clockwise (CCW) beam (\emph{not} roll) 
direction. In the rest frame, since the velocity vanishes, only
Eq.~(\ref{eq:NTLO.1}) influences the vertical motion of the particle
by the Lorentz force,
\begin{equation}
\frac{dp'_y}{dt'}
 = 
-e\gamma(\Delta E_y(\theta) \pm \beta c\Delta B_x(\theta)),
\label{eq:NTLO.3}
\end{equation}
where the sign is appropriate for electrons,
and $\Delta E_y$ and $\Delta B_x$ are unknown lab frame field errors.
Also we are temporarily ignoring the fact that the $\Delta B_x$ term
will have caused CW and CCW closed orbits to differ.
Knowing that the beam stays more or less centered vertically over
long times, averaging this equation over $\theta$ yields the result
\begin{equation}
\langle\Delta E_y\rangle
 = 
\mp \beta c \langle\Delta B_x\rangle.
\label{eq:NTLO.4}
\end{equation}
In particular, if $\langle\Delta B_x\rangle=0$ then $\langle\Delta E_y\rangle=0$.

(This analysis has neglected gravitational forces. There is certainly a vertical 
laboratory gravitational force $(m_e\gamma)g$ acting on each electron 
corresponding to its total
energy $\gamma m_ec^2$. Neglecting the issue
of difference between CW and CCW beams, one can define an ``equivalent''
laboratory magnetic field $\Delta B_x^{\rm G-equiv}$ such that 
\begin{align}
ev\Delta B_x^{\rm G-equiv}
 &=
m_e\gamma g, \notag\\
\quad\hbox{or}\quad
\Delta B_x^{\rm G-equiv}
 &=
\frac{\gamma m_eg}{ec}
 =
5.5\times10^{-18}\,{\rm T}.
\label{eq:NTLO.4p}
\end{align}
Since this is far smaller than the smallest conceivable uncertainty in
the true magnetic field error $\Delta B_x$, I will continue to neglect 
gravity\footnote{Work by Orlov, Flanagan, and Semertzidis 
(which I have not studied) indicates that general 
relativity significantly alters the interpretation of EDM storage
ring experiments. Unlike the purely classical $g=9.8\,{\rm m/s^2}$
gravitational acceleration accounted for here, to be of concern
such an effect must
imply a general relativistic effect capable of mimicking an observable
CP-violating effect. It seems likely to me that any such mechanism
could plausibly be responsible for the observed matter/anti-matter 
imbalance that has motivated the search for non-vanishing EDM's in
the first place. By this reasoning, general relativistic considerations
cannot really alter the motivation for attempting to measure
electric dipole moments. In other words, \emph{if it looks like CP violation, it is CP-violation.}
}.)

We will concentrate, for example, on a beam polarized in the $(y,z)$ plane, 
perpendicular to the local radial direction. The rest frame
polarization vector ${\bf s}$ (conventionally written without a prime in spite
of relating to the rest frame) is given by
\begin{equation}
{\bf s}
 =
{\bf\hat y}\cos\Theta + {\bf\hat z}\sin\Theta.
\label{eq:NTLO.5}
\end{equation}
The error torque acting on the MDM is proportional to ${\bf B}'\times{\bf s}$;
\begin{equation}
{\bf B}'\times{\bf s}
 =
(\Delta B_x(\theta) + \beta\Delta E_y(\theta)/c)\gamma
({\bf\hat z}\cos\Theta - {\bf\hat y}\sin\Theta).
\label{eq:NTLO.6}
\end{equation}
and the design torque acting on the EDM is proportional to ${\bf E}'\times{\bf s}$;
\begin{equation}
{\bf E}'\times{\bf s}
 =
-E\gamma({\bf\hat z}\cos\Theta - {\bf\hat y}\sin\Theta).
\label{eq:NTLO.8}
\end{equation}
Consider first the case of a truly frozen spin beam, meaning $\Theta$ is constant.
The error torque and the design torque are always parallel or antiparallel.
The magnetic error causes a vertical kink in the orbit. Had this kink been
caused by electric field it would have caused no spurious spin precession because
the particle energy is magic. For magnetic bending, the energy is
\emph{not} magic. The resulting precession error is proportional to
the magnetic field error $\Delta B_x$, but the effect is more serious
for the not-very-relativistic magic protons than for the 
fully-relativistic electrons. Comparing Eqs.~(\ref{eq:NTLO.6}) 
and (\ref{eq:NTLO.8}), the discrimination against spurious precession 
can be expressed by an averaged ``EDM selectivity factor'', 
\begin{equation}
{\rm S.F.}
 =
\langle{\rm S.F.}(\theta)\rangle
 = 
\frac{\beta\langle\Delta B_x(\theta)\rangle}{E/c}\,\frac{1}{\gamma_m}.
\label{eq:NTLO.9}
\end{equation}
The $1/\gamma_m$ factor (evaluated at the ``magic'' beam energy)
allows for the fact that it is the difference
between magnetic and electric precession that needs to be 
accounted for. It can be shown, at the magic momentum, that the difference 
between magnetic and electric spin tune precession rates (per radian angular
momentum deflection) is given by
$(d/d\theta)(Q_{\rm MDM,B} - Q_{\rm MDM,e})=1/\gamma_m)$.  Though this
factor is unimportant for the proton EDM measurement, it is important
for the electron measurement, for which $\gamma_m=30$.

Because $\Delta B_x(\theta)$ can have either sign, there will be
significant reduction by averaging of the numerator factor. 
Nevertheless, because the MDM is so huge compared to the EDM, 
in this truly frozen case, this S.F. will be large enough to 
require huge reduction of precession caused by
$\Delta B_x(\theta)$ field errors due to external sources by 
magnetic shielding. 

Consider next the case of rolling polarization beam, meaning 
$\Theta=\omega_{\rm roll}t$ varies sinusoidally. In this case,
replacing $\theta=\omega_0t$, the selectivity factor becomes
\begin{equation}
{\rm S.F.}
 = 
\frac{\beta\langle\Delta B_x(\omega_0t)\cos(\omega_{\rm roll}t)\rangle}{E/c}
\,\frac{1}{\gamma_m}
\approx
0.
\label{eq:NTLO.9p}
\end{equation}
Because the roll frequency is uncorrelated with the rotation frequency, the
average will be essentially zero over times of order the roll period,
say 10\,ms, or longer. What has previously considered to be the most
serious source of systematic error in EDM determination has been
completely eliminated by the rolling polarization. The same cancellation
will occur for all field errors. This ``miracle'' is only brought about
by what is the true miracle, namely the phase-locked, rolling-polarization,
trapped beam. \emph{As noted at the start of this section, this result is incorrect, but has not yet been repaired.}

\subsection{CW/CCW Vertical Beam Separation}
Though the spurious spin precession due to $\Delta B_r$ has been
eliminated, this magnetic field error causes clockwise and
counterclockwise orbits to differ. This magnetic field violates 
the time reversal symmetry of the ring and causes the CW and 
CCW beams to 
separate vertically.  This separation will be limited however
by the opposite electric field components $E_y^{\pm}$ that
the beams encounter because of their different orbits.
Transformed to the electron rest frame, these yield
\begin{align}
{\bf E'}^{(\pm)} 
 &= 
\gamma((\Delta E_y+\Delta E_y^{\pm}){\bf\hat y} \pm \beta{\bf\hat z}\times c\Delta B_x{\bf\hat x})
\notag\\
  &= 
\gamma((\Delta E_y+\Delta E_y^{\pm}) \pm \beta c\Delta B_x){\bf\hat y}
\label{eq:CW-CCW.1} \\
{\bf B'}^{(\pm)} 
 &= 
\gamma(\Delta B_x{\bf\hat x} \mp \beta{\bf\hat z}\times(\Delta E_y+\Delta E_y^{\pm}){\bf y}/c)
\notag\\
  &= 
\gamma(\Delta B_x \pm \beta(\Delta E_y+\Delta E_y^{\pm})/c){\bf\hat x}
\label{eq:CW-CCW.2}
\end{align}
Here $\Delta E_y$ and $\Delta B_x$ are the same lab frame field 
errors as before and $\Delta E_y^{\pm}$ are laboratory frame
electric fields that develop (differently CW and CCW)
to limit the orbit deviations
caused by $\Delta B_x$. Averaged over longitudinal coordinate
$s$ the fields $\Delta E_y^{\pm}$ will be opposite for CW and 
CCW beams but the equality is not guaranteed locally at every
longitudinal position. (This is because the field errors 
do not respect the lattice mirror symmetries.  As it happens, since
the EDM lattice will have only mild beta function
dependence on $s$, the symmetry will be only mildly broken,
causing the fields $\Delta E_y^{\pm}$ to be more or less
equal and opposite locally.)

The rest frame vertical electric fields are
\begin{align}
{E'}_y^{(+)} 
 &= 
\gamma((\Delta E_y+\Delta E_y^{+}) + \beta c\Delta B_x),
\label{eq:CW-CCW.3} \\
{E'}_y^{(-)} 
 &= 
\gamma((\Delta E_y+\Delta E_y^{-}) - \beta c\Delta B_x).
\label{eq:CW-CCW.4}
\end{align}
Upon averaging these yield
\begin{align}
\langle\Delta E_y+\Delta E_y^{+}\rangle = -\langle\beta c\Delta B_x\rangle,
\label{eq:CW-CCW.5} \\
\langle\Delta E_y+\Delta E_y^{-}\rangle = \langle\beta c\Delta B_x)\rangle.
\label{eq:CW-CCW.6}
\end{align}
The rest frame magnetic fields are 
\begin{align}
{B'}_x^{(+)} 
 &= 
\gamma(\Delta B_x + \beta(\Delta E_y+\Delta E_y^{+})/c),
\label{eq:CW-CCW.7} \\
{B'}_x^{(-)} 
 &= 
\gamma(\Delta B_x - \beta(\Delta E_y+\Delta E_y^{-})/c).
\label{eq:CW-CCW.8}
\end{align}
Upon averaging and substituting from Eqs.~(\ref{eq:CW-CCW.5}) and 
(\ref{eq:CW-CCW.6}) these yield
\begin{align}
\langle{B'}_x^{(+)}\rangle 
 &= 
\frac{\langle\Delta B_x\rangle}{\gamma},
\label{eq:CW-CCW.9} \\
\langle{B'}_x^{(-)}\rangle  
 &= 
\frac{\langle\Delta B_x\rangle}{\gamma}.
\label{eq:CW-CCW.10}
\end{align}
The fact that the signs are the same shows that 
the spin precession caused by $\langle\Delta B_x\rangle$ field error 
is the same for CW and CCW beams. The reason for alternating 
between clockwise and counterclockwise circulating beams is
\emph{not}, therefore, to cancel spurious 
$\langle\Delta B_x\rangle$-induced precession. It is to cancel other
sources of systematic error.

If the radial magnetic field causes a net-upward force on an
electron beam then its effect on a counter-circulating electron
beam will be downward. One can therefore measure 
$\langle\Delta B_x\rangle$ by measuring the vertical separation
between counter-circulating beams. 

\subsection{Correcting Orbit Change Caused by $\Delta B_r$}
The vertical deflection at BPM $d$ caused by $N_a$ vertical deflections
$\Delta y'_a$ at lattice positions $a$ is given by\cite{RT-Capri}
\begin{equation}
\frac{y_d}{\sqrt{\beta_d}}
 = 
\sum_{a=1}^{N_a}
\frac{\cos(\mu_y/2 - \phi_{da})}{2\sin(\mu_y/2)}\,
\sqrt{\beta_a}\,\Delta y'_a,
\label{eq:BPM.1}
\end{equation}
where $\mu_y=2\pi Q_y$ is the lattice tune, $\phi_{da}$ is the
vertical betatron phase advance from $a$ to $d$, and $\beta_a$ and 
$\beta_d$ are the corresponding vertical Twiss function values. For the
EDM ring the $\beta$-functions will not depend strongly on 
position. This, simplifies Eq.~(\ref{eq:BPM.1}) to
\begin{equation}
y_d
 = 
\sum_{a=1}^{N_a}
\frac{\cos(\mu_y/2 - \phi_{da})}{2\sin(\mu_y/2)}\,
\beta_a\,\Delta y'_a.
\label{eq:BPM.2}
\end{equation}
For \emph{true frozen spin operation},
to increase rejection of radial magnetic field error $\Delta B_x$,
the vertical tune $Q_y$ would be adjusted intentionally so that
$Q_y<<1$. 
Making the vertical focusing weak amplifies the vertical
separation caused by a given radial magnetic 
field\cite{SemertzidisStrategy}. This
improves the sensitivity with which the field error can be
compensated away by eliminating the beam separation.

The rolling polarization modification has made this unnecessary. 
By allowing vertical tune $Q_y$ to be comparable to horizontal
tune $Q_x$ greatly simplifies the storage ring design. This
will inevitably lead to improved dynamic aperture, and better
suppression of emittance dilution due to intrabeam 
scattering\cite{ValeriIBS}. 

But the rolling polarization brings with it inevitable deflection
errors caused by Wien filter imperfection. This makes it important
to know the orbit shift caused by local $\Delta B_x$ magnetic 
field errors. One of the important strategies for reducing the 
EDM systematic error associated with roll-reversal, is to 
have the roll caused by the reversal of a \emph{single isolated source}.
Nevertheless, for generality, we continue to treat multiple $\Delta B_x$ 
error sources, since precision orbit centering will also be important.
The ultimate EDM accuracy limit will probably concern the accuracy with
which the polarization can be reversed, which will depend on closed
orbit precision. We have been pretending the ring has no field errors, 
which will obviously not be the case. So precise orbit smoothing
will be important. The more accurately the beam can be held on the
design orbit, the more accurate the Wien reversal will be 
(because deviant closed orbit reflects deviant Wien deflection). 

Assuming $N_a$ deflection errors are distributed more
or less uniformly around the ring, an average value for the
cosine factor is $\sin(Q_y\pi)/(Q_y\pi)$. 
The vertical deflection error $\Delta y'_a$ caused by
magnetic field error $\Delta B_{x,a}$ in an element of length
$L_a$ is
\begin{equation}
\Delta y'_a
 =
\frac{c}{p_0c/e}\,\Delta B_{x,a}L_a,
\label{eq:BPM.4}
\end{equation}
and the summation gives approximately 
\begin{equation}
\sum_{a=1}^{N_a}\,\Delta y'_a
 \approx
\frac{c}{p_0c/e}\,
\langle\Delta B_x\rangle\, 2\pi r_0,
\label{eq:BPM.4p}
\end{equation}
Also $\beta_y\approx r_0/Q_y$. 
With these approximations, 
Eq.~(\ref{eq:BPM.2}) becomes 
\begin{equation}
y_d
 \approx
\frac{r^2_0}{Q^2_y}\,
\frac{c}{p_0c/e}\,
\langle\Delta B_x\rangle.
\label{eq:BPM.3}
\end{equation}
The momentum factor can be expressed in terms of $E$ as $p_0c/e=r_0E$,
yielding 
\begin{equation}
y_d
 \approx
\frac{r_0}{Q^2_y}\,
\frac{\langle\Delta B_x\rangle}{E/c}.
\label{eq:BPM.3p}
\end{equation}
The main purpose for the $N_d$ vertical BPM's is to 
accurately measure the average deflection $y_d$ (or
rather, the difference in vertical deflections of
counter-circulating beams) in order to infer  
$\langle\Delta B_x\rangle$.
Defining the precision of each such measurement by
$\sigma_y$, the r.m.s. field error is given by
\begin{equation}
\frac{\sigma_{B_x}}{E/c}
 =
\frac{1}{\sqrt{2N_d}}\,
Q_y^2\,
\frac{\sigma_y}{r_0}.
\label{eq:BPM.5}
\end{equation}
Substituting from Eq.~(\ref{eq:BPM.5}) into 
Eq.~(\ref{eq:NTLO.9}) and using parameter values 
to be spelled out shortly,
\begin{equation}
\frac{\langle\Delta B_x\rangle}{\gamma^2E/c}
 = 
\frac{1}{\gamma^2}
\frac{1}{\sqrt{2N_d}}\,
Q_y^2\,
\frac{\sigma_y}{r_0}
\overset{\rm e.g.}{\ =\ }
0.44\times10^{-13}.
\label{eq:BPM.6}
\end{equation}
For truly frozen spin operation this is the factor to be contrasted 
with the disadvantage factor $\eta_{EM}^{(e)}=0.46\times10^{-15}$ the 
MDM has over an EDM value of $10^{-29}\,$e-cm. 
A priori one does not know $\sigma_y$, but $\sigma_y\approx100\,$nm 
BPM resolution has been achieved, for example in connection with 
the Interanational Linear 
Collider\cite{LightSourcePrec}\cite{LightSourcePrec2}\cite{LiberaBPM}\cite{PrecisionBPM}\cite{EBPM-XBPMcorrelation}. 
This measure suggests the smallest upper bound that can be set on 
the electron EDM will with truly frozen spin operation will be about 
$10^{-27}\,$e-cm. (This is why simultaneous countercirculating beams
were considered to be necessary in the reference\cite{BNLproposal}.)

The same formulas can be used for estimating the precision with 
which closed orbit deviations from the ideal orbit can be minimized
for our actual rolling polarization case. 

\section{Magnetic Shielding}
\subsection{Unimportant Sources of E\&M Noise}
It has been emphasized that the combination of rolling polarization
and phase-locked trap have greatly reduced the sensitivity of the
EDM measurement to electric and magnetic field uncertainties. These
claims are, of course, only valid if successful phase-locked trap 
operation is actually achieved. Magnetic field errors can certainly
prevent phase-locking of the spin wheel and an occasional
magnetic field transient can break the lock, forcing the run in progress
to be aborted. This would be true even assuming perfect polarimetry.

Electromagnetic field transients can also cause runs to be aborted
due to beam loss or emittance dilution, but much 
ordinary storage ring experience exists, 
making further discussion of particle loss unnecessary. 

With electric 
fields of several million volts per meter just about everywhere along the
beam line, the most likely electric transient 
is a spark. Since this could be catastrophic for sensitive electronics 
nearby, this has to be made essentially impossible by appropriate engineering
design. So nothing more will be said about this possibility either. 

Steady, ultralow frequency field errors do not constitute a problem
for phase-locked operation, since the compensation elements can comfortably
cancel their effects. This would include the electric and magnetic fields 
due to miniscule DC leakage currents. It would also include tiny 
patch effect magnetic
fields associated with microscopic mosaic spreads in conductor surface 
structure. The only serious problem to be faced is transient external 
magnetic fields, which are notoriously difficult to shield against.

The polarization roll frequency is autonomously (i.e. self-generated), 
not externally, imposed. With no external source setting the roll frequency, 
no error can result from
the correlation of an external field with the polarization vector. There
could be accidental synchonism. For example the roll frequency could
be 60\,Hz. But this possibility can be recognized and avoided.

\subsection{Low Frequency Magnetic Fields} 
The dominant source of error for all previous EDM experiments 
has come from unknown
spurious spin precession caused by wandering DC (sub-mHz) 
and (less important) low frequency (sub-kHz) magnetic fields remaining in 
spite of all attempts to shield them. 
The differences among competing experimental methods largely come down to
differences in approach to minimization of the errors caused by such
unknown magnetic field variation. Other EDM error sources can only become 
significant once this source has been reduced by several orders of magnitude.

One significant source of man-made noise in this frequency range comes from 
50 or 60\,Hz hum and its harmonics. This noise is sufficiently narrow-banded 
to be avoidable, but only if frequency selectivity is available, by avoiding 
these frequencies. More significant are transients due to starting and stopping 
of electrical machinery, crane and elevator movements, and vehicles passing 
nearby. These noise sources are discussed, for example, in reference~\cite{neutronEDM}, 
and in references given in that paper. Numbers to be used below 
concerning residual magnetic fields are obtained from this reference.

The DC earth magnetic field of about $1\,$mT is not as important
as its r.m.s. variation over typical EDM run durations. After serious 
four layer passive magnetic shielding, the rms error will be about 
$\pm10\,$nT. This five orders of magnitude
reduction is a somewhat ambiguous combination of AC/DC fraction in
the range from 0.01 to 0.1, and a passive permalloy shielding reduction 
ratio in the range from 1000 to 10,000.  Active shielding reduces the 
noise level to about $\pm 1\,$nT. This further reduction of 10 due to 
active field compensation is accurately measureable in any fixed
geometry, but depends strongly on the particular geometry of coils
and experimental apparatus. The effectiveness of active shielding
decreases rapidly with increasing frequency.

\begin{figure}[ht]
\centering
\includegraphics[scale=0.32]{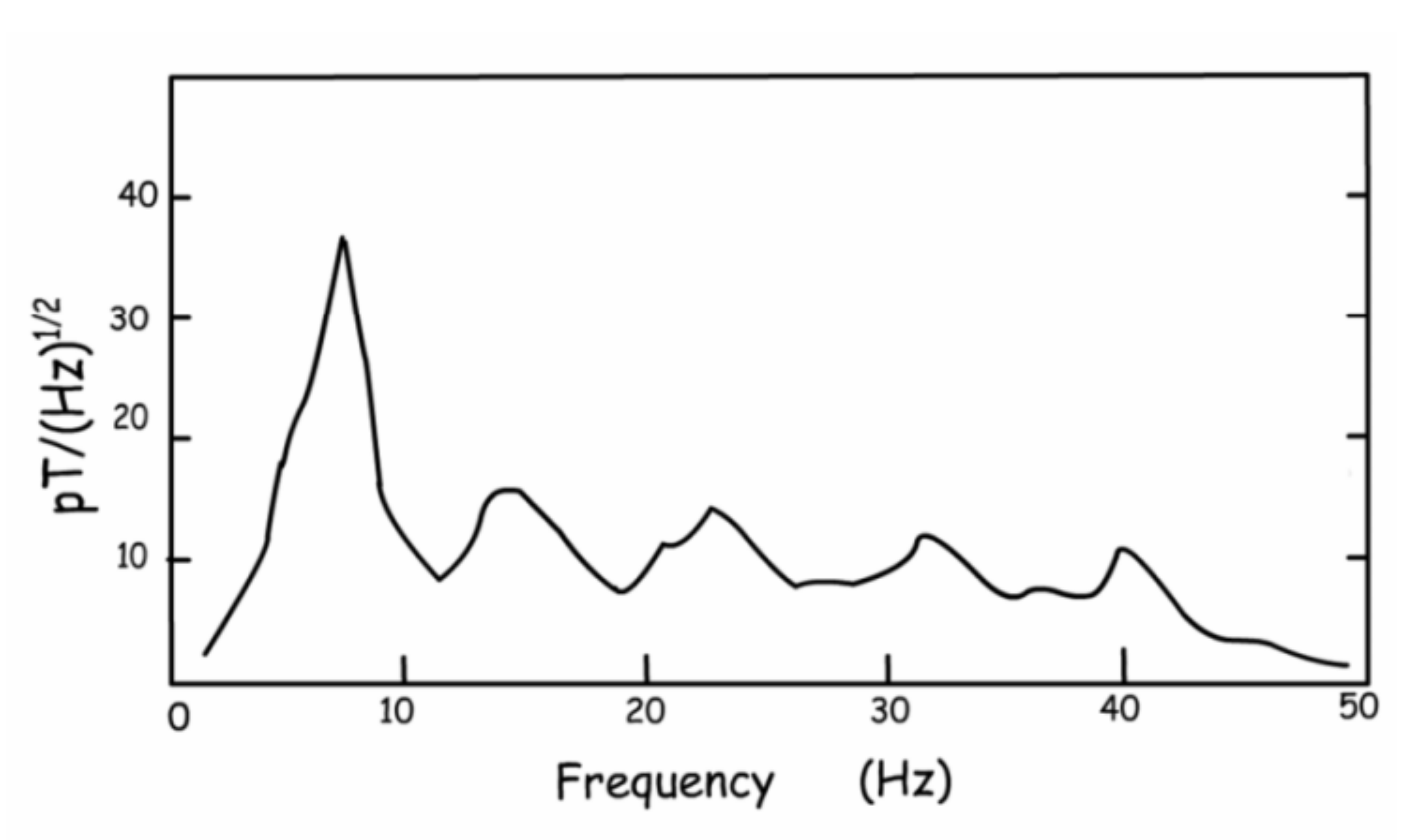}
\caption{\label{fig:TerrestrialB-noise}Natural terrestrial noise in
the frequency range important for rolling polarization EDM 
measurement. Vastly greater noise at ultralow (mHZ) frequencies is
not shown.}
\end{figure}

With resonant polarimetry we can concentrate on higher, but still very low 
frequency, non-man-made, terrestrial, magnetic noise.
According to Bianchi and Meloni\cite{Bianchi-Meloni},
natural terrestrial noise in the frequency range from 1 to 100 Hz 
is dominated by radiation produced in lightning strikes, several million 
per day world-wide and trapped between earth and ionosphere, especially 
at Schumann resonant frequencies, 7.8\,Hz and its lowest few harmonics. 
See Figure~\ref{fig:TerrestrialB-noise} which shows noise levels
of about $\pm 10\,{\rm pT/\sqrt{Hz}}$. Assuming permalloy shielding 
(which continues to work well in this frequency range) an estimated 
residual magnetic field noise in this frequency range, is  
\begin{equation}
\frac{d\sigma_B}{df}[f, {\rm passive\ shielding\ only}] 
\approx 
\pm1\,{\rm fT/\sqrt{Hz}}.
\label{eq:MagNoise.2}
\end{equation}
Active magnetic shielding would be of little value over 
large volumes in this frequency range.

\subsection{Neutron EDM Magnetic Shielding}
The experience with magnetic transients most easily applicable to our
storage ring measurement of charged particle EDM's comes from
the numerous experiments that have been performed to measure the
neutron EDM. A thorough review of all such experiments to find
the most sophisicated, lowest possible residual field procedures, 
is unnecessary. It is more important, at this early stage of 
development of a novel method, to acquire prejudice as to the seriousness 
of the issue, and to estimate the performance to be achieved at 
reasonable cost.
 
In this spirit I review a single recent neutron EDM experimental publication,
chosen, more or less, at random\cite{neutronEDM}, which describes experience
at the Paul Scherrer Institute (PSI). Their shielded volume is roughly a 2.5\,m cube. 
The shielded volume required for our electron EDM experiment would be a
square room, perhaps 7\,m on a side, 3\,m or 4\,m high. A proton EDM experiment 
would perhaps require a shielded 2.5\,m diameter toroidal volume of circumference
300\,m or greater. These dimensions seem close enough for neutron EDM 
experience to be applied to the EDM experiments under discussion.

To obtain neutron EDM measurements to present day
accuracy of roughly $3\times10^{-26}$e-cm accuracy, requires not just the
passive magnetic shielding described previously, but also the further factor of 
10 by active compensation, using compensation coils located outside the 
magnetic shielding, controlled by precision magnetometer 
measurements inside the shield. The PSI investigation found, with 
this \emph{active}
compensation, that ``disturbances 
of the magnetic field are attenuated by factors of 5 to 50 at a bandwidth from 
1\,mHz up to 500\,mHz, which corresponds to integration times 
longer than several hundreds of seconds and represent the important 
timescale for neutron EDM measurement.'' Proton EDM run durations will
be about the same, or somewhat longer.

With passive magnetic shielding at PSI (an accelerator environment)
over times of a few hours, they
find nighttime DC magnetic $B$ field levels changing over a $\pm5\,$nT range
and increasing to $\pm20\,$nT in the daytime. With active shielding 
this is reduced by more than a factor of 10. For round numbers let us say
the residual DC noise, after maximal shielding is at the $1\,$nT level;
\begin{equation}
\frac{\sigma_B}{df}[{\rm DC, passive + active\ shielding}] 
\approx 
\pm 1\,{\rm nT}.
\label{eq:MagNoise.1}
\end{equation}
For the electron and proton EDM measurements proposed here I assume 
similarly high quality passive permalloy magnetic shielding, but
not necessarily active shielding. 
For the electron EDM measurement this would probably
be accomplished by putting the whole experiment in a magnetically shielded room.
This would be impractical for the proton EDM measurement. Rather the ring lattice
would be enclosed in a shielded toroidal tube.
As regards active magnetic compensation, based on PSI experience, I assume 
it would be ineffective and will not be used.

\subsection{Magnetic Noise Comparison of Neutron and Proton EDM Experiments}
To simplify the continuing discussion I will abbreviate the ``Ramsey neutron
EDM'' method as ``nEDM'', the previously proposed storage ring proton EDM 
experiments as ``pEDM-frozen'', and my proposed new proton method 
as ``pEDM-resonant''.

As mentioned earlier, the nEDM, Ramsey separated RF oscillator method, can be
categorized as a frequency domain measurement. In a constant magnetic
field B, the neutron polarization precesses with Larmor frequency proportional
to B. For runs of several hundred second duration one measures an EDM-induced 
phase shift between RF induced polarization induced at the run start, 
from that observed at the end. (The famous Ramsey before and after RF 
pulses improve the precision by converting the center of the fringe pattern 
into a zero crossing.) A spectral pattern is obtained from a 
large number, such as 100, of sequential runs with RF frequency progressively 
advanced by a tiny amount each run. (In this paper I am referring to patterns
like this as fringe patterns.) Repeated 
with electric field off, any non zero EDM would cause a shifted fringe
pattern or, in more literal terms, a frequency shift.  

In the pEDM-resonant method being proposed, the resonant polarimeter monitors 
the precession phase continuously. (This reduces the number of runs from 
100 to 1, but this is not the point here.) The accumulating phase is the sum 
of the EDM phase shift and a (vastly greater) artificially-imposed roll 
phase. The EDM contribution can only be extracted by 
subtracting the result of a subsequent run taken with exactly equal, but 
opposite-sign, artificial roll. In the nEDM method, as already described,
the EDM measurement comes 
from subtracting data taken in sequential runs having different EDM precession 
conditions (such as reversed electric field).
Otherwise, the nEDM method and the pEDM-resonant method
are quite similar. But both are quite different from the previously 
proposed storage ring pEDM-frozen methods. 

As estimated above in Eq.~(\ref{eq:MagNoise.2}), the r.m.s. magnetic field 
uncertainty for the nEDM method, with both active and passive magnetic 
shielding, is $\pm 1\,$nT.
For the pEDM-resonant method the r.m.s. magnetic field uncertainty for
the pEDM-resonant method depends on the resonator bandwidth, which we
take, conservatively, to be 100\,Hz. Using Eq.~(\ref{eq:MagNoise.1})
which assumes passive, but not active shielding, we get the  
the r.m.s. magnetic field uncertainty for the pEDM-resonant method
to be $\pm 10\,$fT.

The nEDM method and the pEDM-resonant methods are sufficiently similar,
as regards sensitivey to unknown magnetic field for the previous two
estimates to be directly comparable. By this comparison
the pEDM-resonant can be expected to be five orders of magnitude more
accurate than the nEDM experiment. Since the quoted
nEDM accuracy is $\pm 3\times10^{-26}$\,e-cm, the pEDM-resonant 
accuracy (coming from this source of error) is estimated to be 
$\pm 3\times10^{-31}$\,e-cm. The fact that this error is so
small almost guarantees that the external magnetic field 
 error source will not, in fact, 
be the dominant error source in the pEDM-resonant method.

(It can be noted, in passing, that this procedure for
estimating the accuracy, 
when applied to the pEDM-frozen method, seems superficially
to imply an accuracy limit in the range from $\pm 3\times10^{-25}$
to $\pm 3\times10^{-26}$\,e-cm,
depending on assumption about the effectiveness of active 
magnetic shielding.
This is because, the random walk of the frozen spins protons is 
just like the random walk of polarized free neutrons (except for 
their minor MDM difference) and is dominated by the wandering
DC magnetic field error.
What makes this inference incorrect is that a different method 
of suppressing magnetic field error is used in 
reference\cite{BNLproposal}. (As already explained) 
measuring and then eliminating
to exquisite accuracy, the relative vertical displacement of 
simultaneously counter-circulating beams, has the effect
of suppressing the wandering DC error in the pEDM-frozen method. 
This is what permitted the pEDM-frozen, 
$10^{-29}$\,e-cm EDM upper limit error estimate.)

As expected, the rolling polarization with resonant polarimetry almost completely 
suppresses all spurious polarization precession caused by magnetic noise. 
Magnetic fields at frequency lower than $f_{\rm roll}$ average to zero after 
multiple rolls.  The effects of magnetic fields at frequency greater than 
$f_{\rm roll}$ average to  essentially zero even during a single polarization roll. 
Only magnetic fields with frequencies close to $f_{\rm roll}$ can produce 
noticeable spurious precession that would emulate an EDM effect. 

There is the possibility of an occasional loss of phase lock due to a 
short magnetic pulse. But the precession impulse angle needed to lose
phase lock is huge compared to the value that would give a seriously
wrong EDM value. During a run a magnetic impulse at this level would
necessarily force the run to be aborted immediately.

\section{Roll Reversal Accuracy}
\subsection{Wien Filter Design}
Because the anomalous precession rates and magic velocities of proton 
and electron are different, their Wien filter designs are not the same.
But, since the essential issues are much the same, only the 
electron case needs to be discussed here. The critical Wien filter, labeled 
$B^W_x$ in Figure~\ref{fig:SingleRing}, causes the electron beam polarization 
to roll with frequency $f_{\rm roll}$, a frequency tentatively in the range
from 10 to 100\,Hz. 

\begin{figure}[ht]
\centering
\includegraphics[scale=0.36]{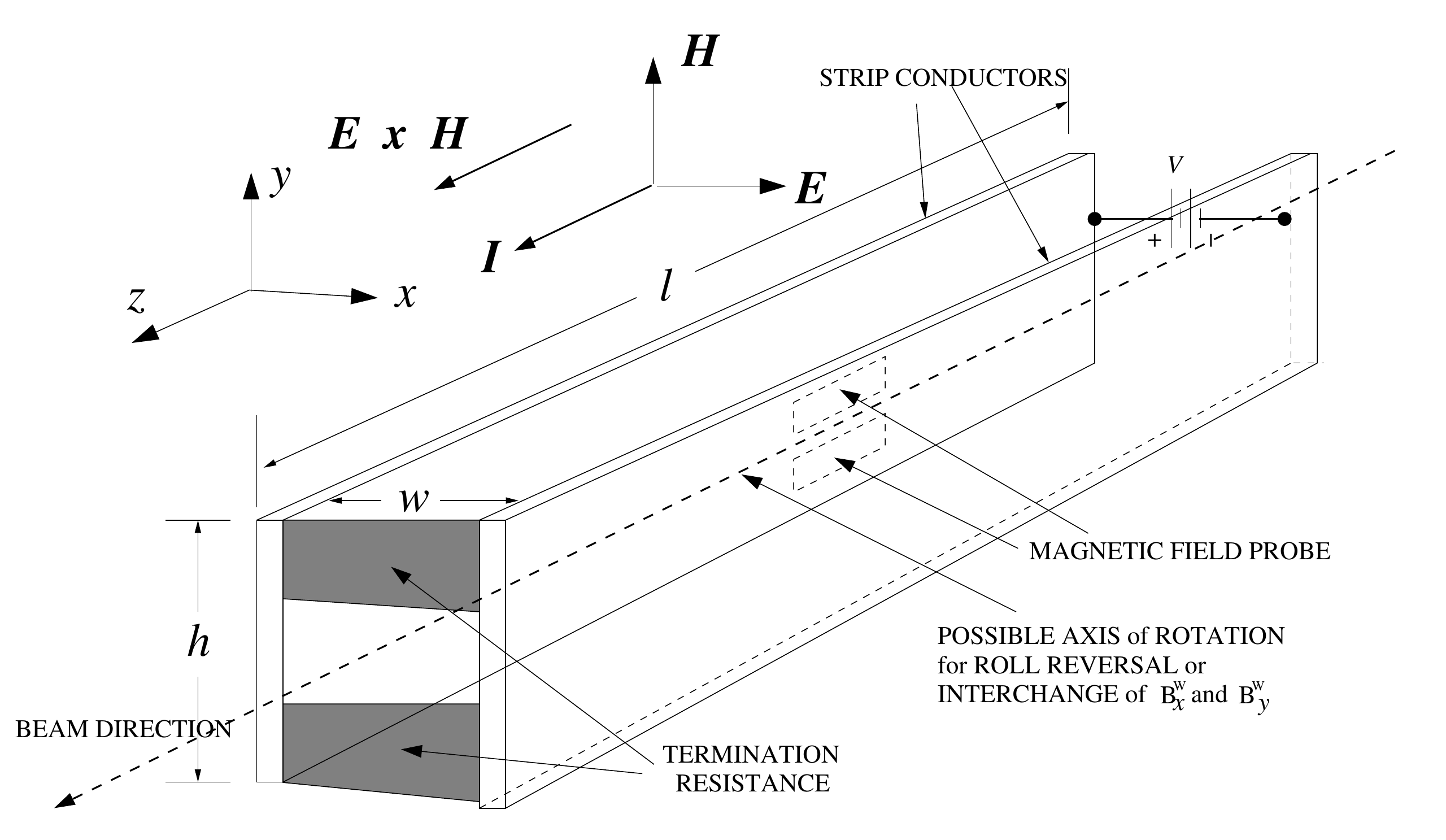}
\caption{\label{fig:StriplineWienFilter}Stripline Wien filter dimensions. With 
electromagnetic power and beam traveling in the same direction, the electric and 
magnetic forces tend to cancel. Termination resistance $R$ is adjusted for 
exact cancelation.}
\end{figure}

A schematic Wien filter design
is shown in Figure~\ref{fig:StriplineWienFilter}. 
Powered from upstream, the Poynting vector ${\bf E}\times{\bf H}$ points
in the beam direction, and the electric and magnetic forces
tend to cancel. With ${\bf E}$ horizontal and ${\bf H}$ vertical, 
using relations~(\ref{eq:Lorentz.4}), the electron rest frame
fields, expressed in terms of laboratory frame quantities, are 
\begin{equation}
E_x' = -\gamma(E_x+\beta c B_y),\quad
B_y' = \frac{B_y}{\gamma}.
\label{eq:Wien.1}
\end{equation}
With Wien filter tuned for zero transverse acceleration in the 
laboratory there is also zero transverse force in the rest frame,
so $E_x'=0$, from which $E_x=-\beta c B_y$. The laboratory frame
time duration within a Wien filter of length $l$ is $l/v$ and
the rest frame duration is $l/(\gamma v)$. While within the Wien
filter in the rest frame, using Eq.~(\ref{eq:Newton.1}), 
the magnetic moment precession angle laboratory advance $\Theta$, 
which is the same as in the rest frame, is given by 
\begin{equation}
\Theta
 =
\frac{1}{\gamma_e^2v_e}\,\frac{g_e\mu_e}{\hbar}\,(B_yl).
\label{eq:Wien.2}
\end{equation}
One of the denominator $\gamma$ factors has come from the second 
of Eqs.~(\ref{eq:Wien.1}), and the other from the time dilation factor.
To produce the desired roll frequency $f_{\rm roll}$ in a ring
with circulation frequency $f_0$, $\Theta$ needs to be given by
\begin{equation}
\Theta
 =
2\pi\,\frac{f_{\rm roll}}{f_0}.
\label{eq:Wien.3}
\end{equation}
In the Wien filter, assuming electrode width $w$ is much greater than
gap $g$, the electric field $E$, the current $I$ and the
magnetic field $H$ are given, in terms of the voltage $V$, and
Wien filter parameters by
\begin{equation}
E=\frac{V}{g},\quad
I=\frac{V}{R},\quad
H=\frac{V}{Rw}.
\label{eq:Wien.3p}
\end{equation}
Then the net transverse force on an electron is
\begin{equation}
F_x= -\frac{eV}{g}\,
\Big( 1 - \frac{v\mu_0g}{Rw} \Big).
\label{eq:Wien.4}
\end{equation}
For exact cancellation
\begin{equation}
R = v\mu_0\frac{g}{w}.
\label{eq:Wien.5}
\end{equation}
(As a check, setting $v=c=1/\sqrt{\mu_0\epsilon_0}$, one obtains 
$R=\sqrt{\mu_0/\epsilon_0}\,g/w$ which is the well known
characteristic impedance of the stripline as a transmission 
line. Even under DC conditions one can describe the setup as a 
wave of infinite wavelength propagating along the line at the speed
of light and being 
perfectly absorbed in the terminating resistor. This is especially
apt for the electron case, since the electron magic velocity is 
close to $c$.)  
Combining formulas, the required length times magnetic-field product 
and current times length product, for electrons, are given by
\begin{align}
Bl &= \frac{v_e\gamma_e^2}{g_e\mu_e/\hbar}\,2\pi\,\frac{f_{\rm roll}}{f_0},\label{eq:Wien.6}\\
Il &= \frac{w}{\mu_0}\,(Bl).
\label{eq:Wien.7}
\end{align}
For $f_0=10\,$MHz, $f_{\rm roll}=100\,$Hz, $g=0.03\,$m, and $w=0.15\,$m, the
termination resistor is $R=(g/w)\times377\,\Omega$, and the  
current times length product is $5.10\,$A-m. 

\begin{figure}[ht]
\centering
\includegraphics[scale=1.0]{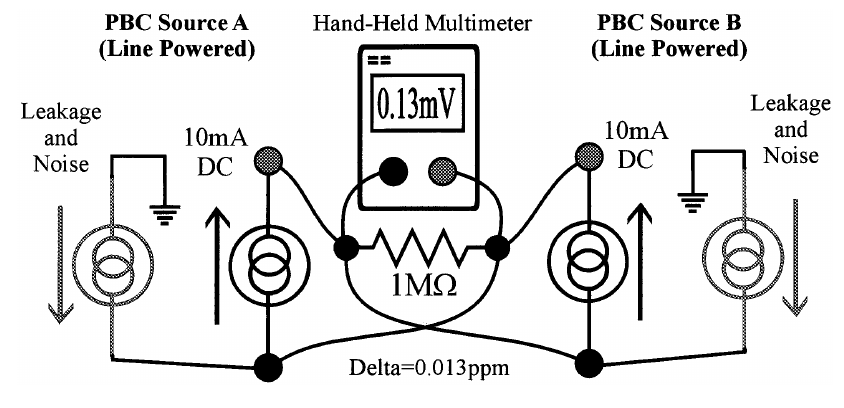}
\caption{\label{fig:CurrentBridge}Current bridge used for high precision current 
monitoring. Copied from CERN PBC reference\cite{CERNpreciseCurrent}. One current can be the
active control current, the other a highly stable reference current. Even hand-held, 1 part
in $10^8$ precision is obtained.}
\end{figure}

Five Amperes is a convenient current for precise control and reversal.
A 5A reference current source from the LHC accerator
control system is DC amplified to produce a $15,000\,A(1\pm2\times10^{-6})$ 
ring magnet supply current\cite{LHCcurrentSupply}. The current calibrator is 
described by Fernquist, Halvarsson, and Pett\cite{CERNcurrentCalibrator}. 
This is, in turn, calibrated from a precise CERN 10\,mA standard current, 
whose specification is year-to-year stability of 1 part in $10^6$ and short 
term reproducibility of 2 parts in $10^7$\cite{CERNpreciseCurrent}. Typical 
variation over 1000 seconds is about 5 parts in $10^8$; its temperature coefficient 
is $2\times10^{-8}/{}^{\circ}\,$K. Meeting these specifications was made
more difficult by the requirement for the unit to be mobile. In a single
shielded location higher precision will be practical. Furthermore, it will
be possible to record every run current to higher precision by comparison
with a standard current, for example using a current bridge circuit like that
copied from reference\cite{CERNpreciseCurrent} and shown in 
Figure~\ref{fig:CurrentBridge}. Run-by-run current reversal accuracy to
several parts in $10^9$ should be achievable in this way. 

Of course the rest of the ring has to be shielded from the magnetic
field caused by these 5\,A currents. Note though, that being a DC magnetic
field, it causes no polarimeter response frequency shift.

Establishing phase-locked, rolling-polarization electron trap operation
will be a tour de force of accelerator technology. But the task of 
setting $B^W_x$ to produce the desired roll frequency $f_{\rm roll}$ to
adequately high accuracy is not very challenging, for example because the
electron MDM is known to such high accuracy. The significant challenge
in measuring the electron EDM to high accuracy will be in guaranteeing
that forward and backward roll frequencies 
(not counting the small EDM effect) are identical to exquisitely high 
accuracy. 

\subsection{Roll Reversal Symmetry}
The basic EDM measurement comes from measuring, say, the forward
roll frequency (as a sideband deviation from the revolution frequency). 
If the MDM-induced roll frequency were perfectly calculable 
(which is far from the case) it could be subtracted off. For 
$f_{\rm roll}=100$\,Hz, and nominal EDM value of $10^{-29}$\,e-cm,
the ratio of imposed roll to EDM roll is about 10 orders of magnitude,
making an absolute subtraction unthinkable. 

On the other hand, both the electron $g_e$-factor and the electron 
MDM $\mu_e$ are known to parts in $10^{13}$. The sideband deviation of 
roll frequency minus revolution frequency can also be measured to high 
accuracy. So, from Eq.~(\ref{eq:Wien.6}), with all factors accurately
known, the $B^W_xl$ product is well known, at least to the extent that 
the electron EDM contribution can be neglected or, more usefully, 
when the electron EDM effect has been arranged to cancel out, irrespective
of its magnitude.

The fundamental measurement comes, therefore, from the difference 
between forward and backward roll frequencies. This transfers
the accuracy requirement to a reversal symmetry requirement. 
The reversal symmetry has to be better than 1 part in $10^{10}$. 
(Optimistically, for the proton EDM measurement, this may be reduceable 
to 1 part in $10^8$.)

The leading task is therefore to minimize the reversal inaccuracy
when the $B^W_x$ Wien filter is reversed. By temperature control the 
Wien dimensions can be controlled to
good accuracy. By interferometry the Wien gap width $g$ can also
be kept quite stable. The magnetic field nearly on axis can also
be measured and held constant to high accuracy.  

To be discussed shortly, the
best measure of the accuracy with which reversal is being done is to
readjust $B^W_x$ and $B^W_y$ to make the polarization roll in the
horizontal, rather than vertical, plane. This completely suppresses
the EDM effect. For best accuracy over 
shortest time interval, this suggest that sets of 4 runs, forward
and backward roll, EDM effect present and suppressed, should
form the basic data collection block. Interleaving these four
configurations repeatedly and in varying order for multiple cycles
may produce smallest systematic error. CW/CCW reversal of beam
direction would then be relatively infrequent, perhaps once per day.

Other reversals are possible---too complicated for thorough analysis here.
Beam rotation can be CW or CCW. Since this requires reversing powered
end and termination end, it is not as clean as one would wish. 
The connections can be switched or, with the apparatus mounted
on a turntable, as in neutron EDM practice, the Wien filter can
be switched end-for-end. Rolling the Wien filters around their axes 
represents another way to cancel reversal bias. 
In combination these reversals can reduce the reversal error. 

Reversing the electric field (as is done in neutron EDM measurement)
is clearly impossible. But, in the electron case, there is the 
possibility of measuring the positron EDM. The electron-positron
EDM difference is probably the most accurate lepton measurement 
possible. 

\subsection{Determination of Roll Reversal Accuracy}
It will be valuable to be able to measure the accuracy with which the 
polarization roll reversal can be accomplished, irrespective of how
good that accuracy may be. One cannot test the reversal accuracy using
an unpolarized beam to suppress the EDM contribution---there would be 
zero resonator response. One can, however, arrange for the polarization
to roll in a horizontal plane. This would force the EDM precession to 
average to zero, while leaving the resonator unchanged. Any residual 
asymmetry could only be ascribed to being instrumental.

To achieve stable horizontal polarization roll will require biasing
the $B^W_y$ Wien filter while leaving $B^W_x$ set for truly frozen 
spin. (This will also entail replacing $B^W_y$ by $B^W_x$ for wheel
stabilization.) In this case the reversal symmetry can be checked with full
digital precision. This may turn out to be the best way to quantify
the reversal symmetry.

Much has been made of the important of the isolation of the $B^W_x$
Wien filter. But, for the reason just given, it will be just as important
for $B^W_y$ to be cleanly isolated from $B^W_x$ and from the rest of the 
ring.

Irrespective of the absolute accuracy of each single roll reversal, the
overall EDM accuracy can be improved by averaging over large numbers
of runs with conditions reversed in ways that should cause asymmetries
to vanish.

\section{Other Calculations}
\subsection{Canceling $\Delta B_z$ Field Errors}
In the proposed ring there will be no
intentional longitudinal field components $B_z$. But there
will be bend field errors $\Delta B_z$. There will therefore
have to be at least one longitudinal trim field $\Delta B_z^{\rm trim}$.  
The rest frame longitudinal fields to be discussed are then
\begin{align}
{\bf B}'
 &=
(\Delta B_z + \Delta B_z^{\rm trim})\,{\bf\hat z}, 
\label{eq:Precession.1} \\
{\bf E}'
 &=
E^{\rm rf}\,{\bf\hat z},
\label{eq:Precession.2}
\end{align}
where $\Delta B_z$ represents an \emph{unknown} longitudinal field component,
and $\Delta B_z^{\rm trim}$ is a \emph{known} longitudinal trim field.
It is unecessary to introduce a time-dependent magnetic RF term, 
because the RF electric field does not interact (on-axis) with the 
MDM. (Off-axis there is a time-dependent magnetic field that has
be discussed separately, along with other issues such as RF 
misalignment.) Various non-zero contributions to 
${\bf E}'$ have also been suppressed, since they give only 
EDM-induced precession small compared to the dominant EDM effect.

Consider a more or less vertically polarized beam, 
with rest frame polarization vector ${\bf s}$ given by
\begin{equation}
{\bf s}
 =
(s_y + \Delta s_y)\, {\bf\hat y} + \Delta s_x\,{\bf\hat x} + \Delta s_z\,{\bf\hat z}.
\label{eq:Precession.3}
\end{equation}
MDM and EDM torques are proportional, respectively, 
to the cross products
\begin{align}
{\bf B}'\times{\bf s}
 &=
(\Delta B_z + \Delta B_z^{\rm trim})\,{\bf\hat z}
\times
\big(( s_y+\Delta s_y)\,{\bf\hat x} - \Delta s_x\,{\bf\hat y}\big)
\notag\\
 &=
(\Delta B_z + \Delta B_z^{\rm trim})
\big(( s_y+\Delta s_y)\,{\bf\hat y} + \Delta s_x\,{\bf\hat x}\big)
\label{eq:Precession.4} \\
{\bf E}'\times{\bf s} 
 &=
E^{\rm rf}\,{\bf\hat z}
\times
\big(( s_y+\Delta s_y)\,{\bf\hat x} - \Delta s_x\,{\bf\hat y}\big)
\notag\\
 &=
E^{\rm rf}\,\big(( s_y+\Delta s_y)\,{\bf\hat y} + \Delta s_x\,{\bf\hat x}\big).
\label{eq:Precession.7}
\end{align}
The only important torque here is magnetic since the electric RF torque
is neglible compared to the bend field torque given in 
Eq.~(\ref{eq:NTLO.8}). The magnetic torque can, potentially, be large
but the leading vertically-directed precession 
$(\Delta B_z + \Delta B_z^{\rm trim})\big(( s_y+\Delta s_y)\,{\bf\hat y}$
will be cancelled by the trim solenoid.
\begin{equation}
\Delta B_z^{\rm trim}
 =
-\langle\Delta B_z\rangle.
\label{eq:Precession.8}
\end{equation}
Since both $\Delta B_z$ and $\Delta B_x$ contribute to unbalancing
the equilibrium, though at right angles, they both have to be
adjusted empirically. If the $B_z$ trimming altered 
${\bf E}'\times{\bf s}$ it would give an unknown EDM effect. 
But there is no such coupling.

\subsection{Geometric Phase Errors}
Spurious precession can result from the failure of commutation
of successive rotations, proportional to the product of the two
precession angles. The MDM precession angles are themselves
so small their squares are negligible. But the artificially
imposed rotation angles are vastly greater. Should a nonzero
EDM value actually be detected, it might need to be corrected
for failure of commutation errors. But this source of error
could not, by itself, produce a statistically significant 
non-zero EDM value of either sign. Any spurious systematic 
phase shift from this source would include the sign of the 
artificial roll and would cancel in the subtraction. 

Geometric phase errors are a significant issue for measuring
the deuteron EDM.
An ideal frozen spin deuteron ring would have electric
and magnetic fields homogeneously superimposed. But segregating the 
electric and magnet fields into disjoint sectors, as suggested
by Senichev\cite{Senichev}, would make the ring construction vastly 
easier to construct and commission.
It has usually been assumed that segregation into electric and magnetic 
sectors in this way would be made unacceptable for EDM measurement
by geometric phase shifts caused by the commutation failure of the
quite large MDM-induced precessions in successive sectors. Cancelation
on the average of the spurious precessions may allow the 
method proposed here to be immune from this systematic error.

\subsection{Small deviations from magic condition}
This section assumes the polarization phase locking described
previously has been successful. With the beam polarization locked 
one can take advantage of the precisely-known electron 
magnetic moment $\mu_e$ and anomalous moment $G_e$ as given
in Table~\ref{tbl:MagneticParams}.
The spin tune $Q^E_s$ relates to precession around the
vertical axis. 
In an all-electric ring $Q_s^E$ is given in
terms of relativistic factor $\gamma$ by
\begin{align}
Q_s^E = G_e\gamma - \frac{G_e+1}{\gamma},
\label{eq:eMDM.2}
\end{align}
At the ``magic'' value, $Q_s^E=0$, $\gamma=\gamma_m$, where
\begin{equation}
\gamma_m = \sqrt{\frac{G_e+1}{G_e}} = 29.38243573.
\label{eq:eMDM.4}
\end{equation}
Solving Eq.~(\ref{eq:eMDM.2}) for $\gamma$, (and requiring $\gamma>0$),
\begin{equation}
\gamma = \frac{Q_s^E + \sqrt{{Q_s^E}^2 + 4G_e(G_e+1)}}{2G_e},
\label{eq:eMDM.5}
\end{equation}
we then obtain
\begin{align}
\Delta\gamma
  &= 
\gamma - \gamma_m  \notag\\
  &= 
\frac{Q_s^E + \sqrt{{Q_s^E}^2 + 4G_e(G_e+1)}}{2G_e} - \sqrt{\frac{G_e+1}{G_e}}.
\label{eq:eMDM.6}
\end{align}
Though this formula is exact, and quite simple, it can usefully be
expanded (to unnecessarily high precision) to
\begin{align}
\gamma
  & =
29.38243573
 + 
431.16379,\frac{\Delta f_y}{f_0}
               \notag \\
  & \qquad + 3163.5\,\left(\frac{\Delta f_y}{f_0}\right)^2
 + \cdots.
\label{eq:eMDM.7}
\end{align}
Here $Q^E_s$ has been re-expressed in terms of the frequency
deviation from magic, $\Delta f_y=f_y-f_m$, of the polarization
around a vertical axis.
This formula is intended for use only near $\gamma_m$, with
the ratio $\Delta f_y/f_0$ being a tiny number, 
less than $10^{-5}$ for example.
Further terms are easily obtained, but for values of $\Delta\gamma$
large enough to require them one can simply use 
Eq.~(\ref{eq:eMDM.6}). With the RF frequency
known to arbitrarily high accuracy and the beam polarization
locked to the revolution frequency, measuring the polarimeter
response frequency establishes $\gamma$ to correspondingly high
accuracy. 

One is accustomed to expecting a spread of spin tunes 
due to finite betatron and synchrotron amplitudes and energy 
deviation of each individual particle from the central beam 
$\gamma$ value. In a certain sense the beam conditions are 
better defined than this. 
As long as decoherence can be neglected (i.e. for times short 
compared to SCT) the net angular precession of each spin vector
cannot exceed $\pi/2$. During this time $\gamma=\gamma_m$ 
for every particle on the average. One sees that phase locking 
of the polarization establishes the mean beam parameters to exquisitely 
high accuracy. Of course this remarkable behavior begins to break down for 
times approaching SCT. In effect, loss of phase lock and decoherence 
of the beam polarization are two manifestations of the same spin evolution. 
Achieving large SCT and maintaining the possibility of phase locking the 
polarization are equivalent tasks.

This analysis has assumed a purely electric lattice. In practice
there will be some average vertical magnetic field component
$\langle B_y\rangle$ that will cause the $\gamma$ value determined by
Eq.~(\ref{eq:eMDM.7}) to be not quite correct.
But it has also been anticipated that keeping the polarimeter loop
locked will require a Wien filter which compensates for the
(unknown) total angular deflection $\Delta\theta_{\rm error}$ 
by magnetic fields. The spin tune ascribable to this lattice
modification is 
\begin{align}
Q_s^{W(\Delta\theta)}
 &=
\frac{\Delta\theta_{\rm error}}{2\pi}\,
\left(G_e\gamma - G_e\gamma +\frac{G_e+1}{\gamma}\right)  \notag\\
 &\approx
\frac{\Delta\theta}{2\pi}\,\frac{G_e+1}{\gamma_m} 
 =
\frac{\Delta\theta_{\rm error}}{2\pi}\,\frac{1.00116}{29.38243}.
\label{eq:eMDM.8}
\end{align}
which compensates for fraction $\Delta\theta_{\rm error}/(2\pi)$ 
of each turn being caused by magnetic rather than
electric field. The (well, but not perfectly, known)
strength $I^W_x$ of the Wien filter 
excitation can be interpreted as a measurement of the angular
bend caused by unknown magnetic field component 
$\langle B_y\rangle$ averaged over the ring. The ultimate
claimed $\gamma$ precision depends on this correction.
In the actual experiment, since the roll is around the
radial axis, the spin tune will necessarily be exactly zero,
meaning the phase locking requires the Wien filter 
to exactly cancel $\langle B_y\rangle$.

One problematical phase-locked loop detail concerns the box labelled
``polarity reversal'' in Figure~\ref{fig:SingleRing}. It would
be splendid if the roll reversal could be completed without losing
phase lock. But reversing the roll introduces a discontinuous phase 
shift in the polarization signal inputs to the resonant polarimeters 
which will force the 
circuits to ring down and then ring up again, unlocking the closed
loop synchronization. 

Perhaps the roll drive phase could be judiciously reversed at just 
the right time to avoid this effect and preserve 
all phase locking? If this could be done then the roll orientation could 
be reversed at a quite high rate within each fill cycle, greatly reducing 
systematic frequency shift errors. If not, then roll reversal may have
to be limited to once per fill. However there is an excellent reason for
\emph{not} attempting to synchronize any control function with
the roll frequency---it would contradict the claim made earlier in
the paper that no control sources are synchronous with the roll
frequency. Violating this seems sure to cause trouble. This seems to
guarantee loss of phase-lock during roll reversal.

\section{Recapitulation and Conclusions}
The purpose of the paper has been to describe a method for measuring
EDM's of electrons and protons with far higher accuracy than they are
presently known. The essential change from earlier proposals comes
about from the introduction of resonant polarimetry, which permits 
phase-locked loop, rolling-polarization trap operation, and shifts the
EDM signal into the frequency domain. Important contentions, conclusions
and conjectures are included in the following long list:
\begin{itemize}
\item
Successful application of the method depends on two
not yet established experimental methods: resonant polarimetry (promising
theoretically) and ``rolling polarization trap'' operation---meaning
stable, phase-locked, rolling polarization operation---(promising experimentally). 
\item
It is thermal noise in the resonant polarimeter that limits EDM
measurement \emph{precision}.
\item
A successful single beam fill will include at least one forward/backward
reversal of the roll (not beam) direction,  with roll frequency precisely 
measured both before and after. This will produce one \emph{precise} EDM 
measurement.  
\item 
It is frequency domain, digital, frequency scaling, that enables at least one
reversal per fill, giving one precise EDM measurement for every roll reversal.
\item
A successful single beam run will consist of four data sets,
EDM effect on and off, roll forward and backward.
\item
CW/CCW beam direction reversal will reduce systematic error.
\item
Expressed as EDM upper limit, measurement \emph{precision} 
of $10^{-30}\,$e-cm after year-long running, for either electrons and protons,
can be expected.
\item
\emph{Accuracy} at the same level as the \emph{precision} will require average sign
reversal accuracy of Wien filter length/strength product 
at the level of one part in $10^{11}$, also averaged over one year.
Lower reversal nulling accuracy will give proportionally lower EDM
measurement accuracy. 
\item
Apparatus constituting a single Wien filter will all be
contained in a single, temperature regulated, limited vibration, magnetically
shielded, highly isolated, etc. box.
\item
``Rolling polarization trap'' operation greatly improves EDM selectivity
(essentially eliminating spurious external field induced precession) and 
permits the rest of the ring to have relaxed requirements. 
\item
Identical beam distribution requirement for oppositely-directed beams are
similarly relaxed by the requirement of precise EDM measurement for every fill.
\item
The greatly reduced sensitivity to unknown radial magnetic field eliminates
the need for counter-circulating beams and ultra-low vertical tune $Q_y$.
This, in turn, allows relatively strong alternating gradient ring design,
providing larger aperture and higher stored charge.
\item
Emittance growth due to intrabeam scattering (IBS) has been seen as a serious
impediment to EDM measurement. With ultra-low $Q_y$ not required, 
running ``below transition'', which tends to suppress beam growth due to IBS, 
will be possible.
\item
Electron cooling has proved (at Juelich) to be highly effective in increasing
SCT. Though long SCT would improve EDM accuracy, electron cooling has 
previously been rejected out of hand, because of its uncontrolled electric and 
magnetic field
errors it would introduce. Rolling polarization may remove this impediment
to the use of electron cooling.
\item
Spin coherence time SCT is greatly increased by M\"obius ring operation.
\end{itemize}
Recapitulating the recapitulation, all of these wonderful EDM measurement
improvements have been brought about by the phase-locked loop, storage ring
polarized-beam trap, made possible by resonant polarimeter. These
advances have not yet been achieved in practice. The most immediate task is
to successfully build and test resonant polarimetry. 

I have profited especially from conversations with Bill Morse, Yuri Orlov,
Frank Rathmann, Yannis Semertzidis, and John Talman, as well as with 
Jim Alexander, Mei Bai, Ivan Bazarov, Mike Blaskowitz, Peter Cameron, Bruce Dunham, 
Ralf Gebel, Tom Kinoshita, Andreas Lehrach, Alfredo Luccio, Nikolay Malitsky, 
Bob Meller, Maxim Perelstein, Michael Peskin, Thomas Roser, Dave Rubin, Anders Ryd, 
Yuri Senichev, Valeri Shemelin, Eric Smith, Ed Stephenson, and Hans Stroeher. 
And others.

\end{document}